\documentclass[prd,twocolumn,tshowpacs,preprintnumbers,amsmath,nofootinbib,amssymb,floatfix]{revtex4-1}
\usepackage{graphicx}
\usepackage[sort&compress]{natbib}
\usepackage{subfigure}
\usepackage{bm}
\usepackage{amsfonts}
\usepackage{cancel}
\usepackage{lmodern,dsfont}

\begin{document}
\arraycolsep1.5pt
\newcommand{\Ima}{\textrm{Im}}
\newcommand{\Rea}{\textrm{Re}}
\newcommand{\mev}{\textrm{ MeV}}
\newcommand{\be}{\begin{equation}}
\newcommand{\ee}{\end{equation}}
\newcommand{\bea}{\begin{eqnarray}}
\newcommand{\eea}{\end{eqnarray}}
\newcommand{\bef}{\begin{figure}}
\newcommand{\eef}{\end{figure}}
\newcommand{\ba}{\begin{eqnarray}}
\newcommand{\ea}{\end{eqnarray}}
\newcommand{\gev}{\textrm{ GeV}}
\newcommand{\nn}{{\nonumber}}
\newcommand{\dtres}{d^{\hspace{0.1mm} 3}\hspace{-0.5mm}}
\newcommand{\rts}{ \sqrt s}
\newcommand{\non}{\nonumber \\[2mm]}

\newcommand{\re}{\text{Re }}
\newcommand{\im}{\text{Im }}
\newcommand{\bra}[1]{\langle \, #1 \, |}
\newcommand{\ket}[1]{| \, #1 \, \rangle}

\title{A narrow $DNN$ quasi-bound state}

\author{M. Bayar$^{1,2}$, C. W. Xiao$^1$, T. Hyodo$^3$, A. Dot\'e$^{4,5}$, M. Oka$^{3,5}$ and E. Oset$^1$}
\affiliation{
$^1$Departamento de F\'{\i}sica Te\'orica and IFIC, Centro Mixto Universidad de Valencia-CSIC,
Institutos de Investigaci\'on de Paterna, Aptdo. 22085, 46071 Valencia,
Spain \\
$^2$Department of Physics, Kocaeli University, 41380 Izmit, Turkey \\
$^3$Department of Physics, Tokyo Institute of Technology, 
Meguro 152-8551, Japan \\
$^4$High Energy Accelerator Research Organization (IPNS/KEK),
1-1 Ooho, Tsukuba, Ibaraki, Japan, 305-0801 \\
$^5$J-PARC Branch, KEK Theory Center, Institute of Particle and Nuclear Studies, High Energy Accelerator Research Organization (KEK), 203-1, Shirakata, Tokai, Ibaraki, 319-1106, Japan
}

\date{\today}

\begin{abstract}

The energies and widths of $DNN$ quasi-bound states with isospin $I=1/2$ are evaluated in two methods, the fixed center approximation to the Faddeev equation and the variational method approach to the effective one-channel Hamiltonian. The $DN$ interactions are constructed so that they dynamically generate the $\Lambda_c(2595)$ ($I=0$, $J^{\pi} =1/2^-$) resonance state. We find that the system is bound by about 250 MeV from the $DNN$ threshold, $\sqrt{s} \sim 3500$ MeV. Its width including both the mesonic decay and the $D$ absorption, is estimated to be about $20$-$40$ MeV. The $I=0$ $DN$ pair in the $DNN$ system is found to form a cluster that is similar to the $\Lambda_c(2595)$.

\end{abstract}

\maketitle

\section{Introduction}
\label{Intro}

The interaction of mesons with nuclei and the property of mesonic bound states are one of the most important topics in the nuclear-hadron physics~\cite{Krell:1969xn,Toki:1989wq,Nieves:1993ev,Batty:1997zp,okumura,Hayano:2008vn}. Bound states of pions and $K^-$ have been investigated for long and have revealed the roles of strong interactions in the hadron-nucleus bound states. A step forward in the experimental observation of the most deeply bound pionic states was given using the $(d,^{3}$He) reaction \cite{Hirenzaki:1991us}, and also, although less clearly, using the coherent radiative $\pi^-$ capture~\cite{Nieves:1991yf} in Ref.~\cite{Raywood:1997vu}. The deeply bound kaon atoms had been studied theoretically using the optical potentials~\cite{Friedman:1994hx,Lutz:1997wt,angelsnuc,SchaffnerBielich:1999cp,Cieply:2001yg}. Because of the large imaginary part, the width of the bound states is larger than the energy separation between the levels~\cite{okumura,baca}, so that the experimental observation is not feasible (see also Refs.~\cite{Magas:2006fn,Oset:2007vu}).

The simplest of the many-body kaonic nuclear system is the $\bar{K}NN$, which has had much attention theoretically. Because the $\Lambda(1405)$ resonance is interpreted as a quasi-bound state of the $\bar{K}N$ system in the $\pi\Sigma$ continuum~\cite{Dalitz,Kaiser:1995eg,Oset:1998it,Oller:2000fj,Lutz:2001yb,Jido:2003cb,Hyodo:2011ur,IHW}, one expects a quasi-bound $\bar{K}NN$ system driven by the attractive $\bar{K}N$ interaction in the isospin $I=0$ channel. Various approaches have resulted in a rather general consensus that the quasi-bound state is obtained above the $\pi\Sigma N$ threshold and the width is larger than the binding~\cite{Ikeda:2007nz,Shevchenko:2006xy,Shevchenko:2007zz,Dote:2008in,Dote:2008hw,Ikeda:2008ub,akayama,Ikeda:2010tk,melahat,Oset:2012gi,Bayar:2012rk,Barnea:2012qa}. Thus, the experimental identification of this system would be difficult.

What we report here is the analogous state of the $\bar{K}NN$, substituting the $\bar{K}$ by a $D$ meson. The $D N$ interaction in $I=0$ is predicted to be attractive in the vector meson exchange picture, and thus to dynamically generate the $J^{P}=1/2^{-}$ excited state, $\Lambda_c(2595)$ \cite{Hofmann:2005sw,mizuangels,Tolos:2007vh,GarciaRecio:2008dp}. The $\Lambda_c(2595)$ resonance is rather narrow ($\Gamma < 1.9 ~$MeV), in contrast to the analogous $\Lambda(1405)$ with apparent widths of the order of $30$-60 MeV \cite{Jido:2003cb,Hyodo:2011ur,IHW}. While the large width of the $\Lambda(1405)$ is responsible for the large width of the  $\bar{K}NN$ state, the analogous state $DNN$, where the $\Lambda_c(2595)$ plays the role of the $\Lambda(1405)$ in the  $\bar{K}NN$ state, has much better chances to survive as a long lived and observable state.
    
The interaction of the $D$ mesons with nuclei has been addressed in Refs.~\cite{mizuangels,Tolos:2009nn,GarciaRecio:2008dp} and the possibility of making bound atomic states of $D$ mesons in nuclei has been considered in Ref.~\cite{GarciaRecio:2010vt}. The interaction of $DN$ is attractive both in isospin $I=0$ and $I=1$, but much weaker for $I=1$. Although this leads to a weakly attractive $D^+p$ interaction, the Coulomb repulsion becomes important for heavier nuclei. As a consequence, the $D$ will be only weakly bound in heavy nuclei and the probability to see these bound states is not excessively promising \cite{GarciaRecio:2010vt}. However, few-body systems like $DNN$ are less affected by the Coulomb repulsion particularly for the total isospin $I_{\text{tot}}=1/2$. 

With this in mind we tackle the $DNN$ system from two different approaches. The first one is using the fixed center approximation (FCA) to the Faddeev equations, as done in Refs.~\cite{melahat,Oset:2012gi,Bayar:2012rk} for the $\bar{K}NN$ system. The second one is using the variational method as done in Refs.~\cite{Dote:2008in,Dote:2008hw}. In order to gain confidence that the state found is narrow, we have also evaluated the width of the state coming from the absorption of the $D$ by a pair of nucleons going to the $\Lambda_c N$ system in the FCA, analogous to the absorption of $\bar{K}$ by a pair of nucleons as considered in Refs.~\cite{angelsnuc,Dote:2008hw,Sekihara:2009yk}. In the variational approach, we extract typical size of the quasi-bound state from the obtained wave function. 
    
The paper is organized as follows. In Sec.~\ref{sec:DN}, we briefly introduce the coupled-channel approach for the $DN$ scattering and derive the corresponding $DN$ potential. These provide the basis of the three-body calculations in later sections. The FCA to the Faddeev equations is formulated in Sec.~\ref{sec:FCA}, together with the evaluation of the two-nucleon absorption. The variational approach to the same $DNN$ system is discussed in Sec.~\ref{sec:variational}. The numerical results of the three-body calculations are shown in Sec.~\ref{sec:results}. The discussion for the obtained results are given in Sec.~\ref{sec:discussion}. The conclusions of this study are drawn in the last section.

\section{$DN$ scattering and interaction}\label{sec:DN}

We consider the two-body $DN$ scattering based on the model in Ref.~\cite{mizuangels}. This is a coupled-channel approach to the $s$-wave meson-baryon scattering in the vector-meson exchange picture. The negative parity $\Lambda_{c}(2595)$ resonance is dynamically generated as a quasi-bound state of the $DN$ system in the $I=0$ channel, just like the $\Lambda(1405)$ resonance in the strangeness sector~\cite{Kaiser:1995eg,Oset:1998it,Oller:2000fj,Lutz:2001yb,Jido:2003cb,Hyodo:2011ur,IHW}. In Sec.~\ref{subsec:amplitude}, we derive the $DN$ two-body scattering amplitude which will be used in the FCA calculation. An effective single-channel potential is constructed so as to reproduce the equivalent scattering amplitude in section~\ref{subsec:potential}. This will be the basic input in the variational calculation. We work in the isospin symmetric limit, which is sufficient for the required precision of the present study.

\subsection{Coupled-channel model for the $DN$ scattering}\label{subsec:amplitude}

\begin{figure}[tbp]
    \centering
    \includegraphics[width=0.5\textwidth]{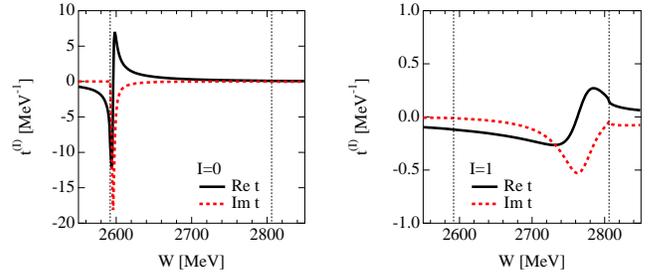}
    \caption{\label{fig:amplitude}
    (Color online) $S$-wave $DN$ scattering amplitude in the coupled-channel model~\eqref{eq:Tamp} (Left: $I=0$ channel, Right: $I=1$ channel). Vertical dotted lines represent the threshold energies of $\pi\Sigma_{c}$ and $DN$ channels.
    }
\end{figure}

We consider seven (eight) coupled channels in the isospin $I=0$ ($I=1$) sector, $DN$, $\pi\Sigma_{c}$, $\eta\Lambda_{c}$, $K\Xi_{c}$, $K\Xi_{c}^{\prime}$, $D_{s}\Lambda$, and $\eta^{\prime}\Lambda_{c}$ ($DN$, $\pi\Lambda_{c}$, $\pi\Sigma_{c}$, $\eta\Sigma_{c}$, $K\Xi_{c}$, $K\Xi_{c}^{\prime}$, $D_{s}\Sigma$, and $\eta^{\prime}\Sigma_{c}$). In Ref.~\cite{mizuangels}, the coupled-channel interaction is given by the Weinberg-Tomozawa term
\begin{equation}
    v_{i j}^{(I)}(W) 
    = -  \frac{\kappa C_{i j}^{(I)}}{4 f^2}
    (2W - M_{i}-M_{j}) 
    \sqrt{\frac{M_i+E_i}{2M_i}}\sqrt{\frac{M_j+E_j}{2M_j}} ,
    \nonumber
\end{equation}
where $W$ is the total energy, $f$ is the meson decay constant, $M_{i}$ and $E_{i}$ are the mass and energy of the baryon in channel $i$, respectively,  and $C_{ij}^{(I)}$ is the group theoretical coupling strength for isospin $I$. The reduction factor $\kappa$ is introduced to take into account the mass difference of the exchanged meson, which we set $\kappa=1$ ($\kappa=\kappa_{c}=1/4$) for the $uds$ (charm) flavor exchange process~\cite{mizuangels}. The scattering amplitude $t_{ij}$ is obtained from the matrix equation
\begin{equation}
   t^{(I)} = ((v^{(I)})^{-1}-g^{(I)})^{-1}
   \label{eq:Tamp} ,
\end{equation}
where the diagonal loop function is given in dimensional regularization as
\begin{align}
  &g_i^{(I)}(W;a_i(\mu)) \nonumber \\
  =
  &\frac{1}{(4\pi)^{2}}
    \Bigl\{a_i(\mu)+\ln\frac{M_i^{2}}{\mu^{2}}
    +\frac{m_i^{2}-M_i^{2}+W^{2}}{2W^{2}}
    \ln\frac{m_i^{2}}{M_i^{2}}
    \nonumber\\
    &+\frac{\bar{q}_i}{W}
    [\ln(W^{2}-(M_i^{2}-m_i^{2})+2W\bar{q}_i)
    \nonumber\\
    &+\ln(W^{2}+(M_i^{2}-m_i^{2})+2W\bar{q}_i) 
    \nonumber\\
    &
    -\ln(-W^{2}+(M_i^{2}-m_i^{2})+2W\bar{q}_i)
    \nonumber\\
    &-\ln(-W^{2}-(M_i^{2}-m_i^{2})+2W\bar{q}_i)
    ]\Bigr\} ,  
    \nonumber
\end{align}
where $\bar{q}_{i}$ is the magnitude of the three-momentum in the center-of-mass frame. We choose the subtraction constants at $\mu=1$ GeV as 
\begin{align}
  a_{DN} 
  =& -2.056
  ,  
  \quad 
  a_{i} = -2.06 \ (i\neq DN) ,
  \label{eq:subtraction}
\end{align}
for both the $I=0$ and $I=1$ states, so that the $\Lambda_{c}(2595)$ resonance  is dynamically generated at the observed energy. By choosing the isospin symmetric subtraction constants~\eqref{eq:subtraction}, a resonance state is also generated in $I=1$ at $\sim 2760$ MeV. The diagonal components of the $s$-wave scattering amplitudes in the $DN$ channel, which are complex above the $\pi Y_{c}$ ($Y_{c}=\Lambda_{c}, \Sigma_{c}$) threshold, are shown in Fig.~\ref{fig:amplitude}. The resonant nature of the amplitudes can be seen in both channels.

It is worth comparing the $I=0$ $DN$-$\pi\Sigma_{c}$ system with the corresponding $\bar{K}N$-$\pi\Sigma$ system. Both the systems have a quasi-bound state. Neglecting the small effect of the normalization factor, we can write the coupling strength for the $DN$ case as
\begin{align}
    v_{ij}
    \sim & 
    \begin{pmatrix}
    3 & \sqrt{\frac{3}{2}}\kappa_{c} \\
    \sqrt{\frac{3}{2}}\kappa_{c} & 4 \\
    \end{pmatrix} \frac{2W - M_{i}-M_{j}}{4f^{2}} ,
    \nonumber
\end{align} 
where the channels are assigned as $DN$ ($i=1$) and $\pi\Sigma_{c}$ ($i=2$). This is the same form with the $\bar{K}N$-$\pi\Sigma$ case, except for the factor $\kappa_{c}=1/4$ in the off-diagonal channel. The diagonal interaction is proportional to the meson energy $W-M_{i}$, which is reduced to the meson mass at threshold. Thus, there are three differences from the strangeness sector: 1) heavy mass of $D$ meson, which enhances the strength of the $DN$ interaction by the energy factor $W-M_{i}$; 2) large reduced mass of the system, which suppresses the kinetic energy in the charm sector; 3) weak transition coupling $DN\to \pi\Sigma_{c}$, which suppresses the decay of the quasi-bound state into the $\pi\Sigma_{c}$ state. These facts explain the reason why the $DN$ quasi-bound state is generated with larger binding energy and narrower width than those of the $\bar{K}N$ quasi-bound state. In addition, 1) and 2) also enhance the attractive interaction in the $I=1$ channel. As a consequence, we obtain a resonance state also in $I=1$ at $\sim 2760$ MeV, as far as choosing the isospin symmetric subtraction constants~\eqref{eq:subtraction}. 

\subsection{Effective single-channel $DN$ potential}\label{subsec:potential}

Now we construct an effective single-channel potential, which will be used in the variational calculation of the $DNN$ system. We utilize the method in Ref.~\cite{Hyodo:2007jq}, first constructing a single-channel framework which is equivalent to Eq.~\eqref{eq:Tamp} and then translating the result into a local and energy-dependent potential in coordinate space. 

The effective interaction $v^{\text{eff}}$ is constructed to reproduce the original amplitude $t_{11}$, given by the $DN$ single-channel scattering equation (we suppress the isospin index in this section)
\begin{align}
    t_{11}
    =& [(v^{\text{eff}})^{-1}-g_1]^{-1} .
    \label{eq:singlechannel}
\end{align}
It is shown that the $v^{\text{eff}}$ is given by the sum of the bare interaction in channel 1 $(v_{11})$ and the term with coupled-channel effects as~\cite{Hyodo:2007jq}
\begin{align}
    v^{\text{eff}}
    =&v_{11} 
    + 
    \sum_{m=2}^{N} v_{1m} g_m 
    v_{m1}
    + 
    \sum_{m,l=2}^{N} v_{1m} g_m t^{(N-1)}_{ml}g_l 
    v_{l1}
    \label{eq:Veffective} , 
\end{align}
where $t^{(N-1)}_{ml}=[(v^{(N-1)})]^{-1}-g^{(N-1)}]^{-1}$ is the $(N-1)\times (N-1)$ matrix of the coupled-channel amplitude without the $DN$ channel. In this way, Eq.~\eqref{eq:singlechannel} gives the equivalent amplitude with the $11$ component in Eq.~\eqref{eq:Tamp}. $v^{\text{eff}}$ is complex above the $\pi Y_{c}$ threshold, because of the imaginary part of the loop function of the $\pi Y_{c}$ channel in Eq.~\eqref{eq:Veffective}.

\begin{figure}[tbp]
    \centering
    \includegraphics[width=0.5\textwidth]{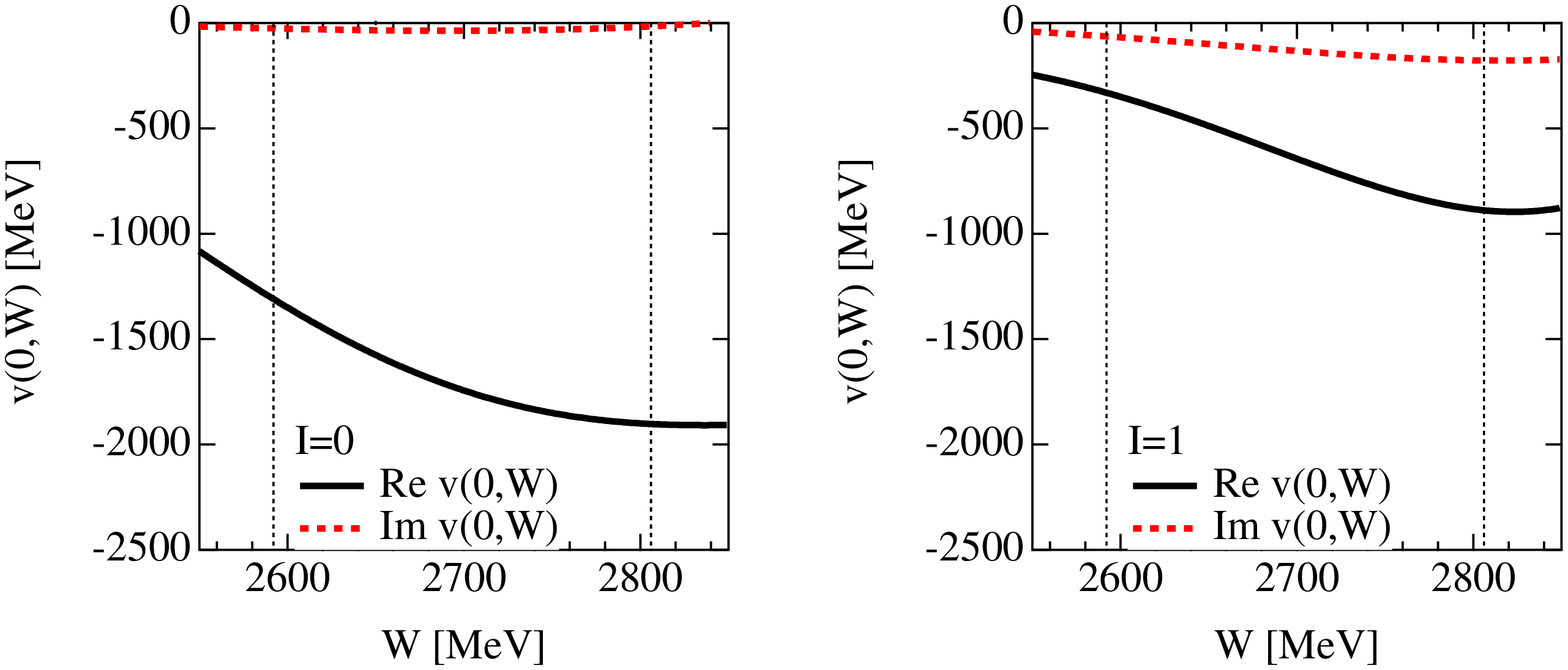}
    \caption{\label{fig:potential}
    (Color online) Strength of the effective potential $v_{DN}(r,W)$ at $r=0$ by Eq.~\eqref{eq:DNpotential} (Left: $I=0$ channel, Right: $I=1$ channel). The range parameter is chosen to be $a_{s}=0.4$ fm. Vertical dotted lines represent the threshold energies of $\pi\Sigma_{c}$ and $DN$ channels.}
\end{figure}

We then translate $v^{\text{eff}}$ into the local potential in coordinate space. Adopting a single gaussian form for the spatial distribution, the two-body potential can be written as
\begin{align}
    v_{DN}(r;W)
    =&
    \frac{M_{N}}{2\pi^{3/2}a_{s}^{3}\tilde{\omega}(W)} 
    \nonumber \\
    &\times 
    [v^{\text{eff}}(W)
    +\Delta v(W)]
    \exp[-(r/a_{s})^{2}]  , \quad
    \label{eq:DNpotential}
\end{align}
where $a_{s}=0.4$ fm is the range parameter of the potential and $\tilde{\omega}(W)$ is the reduced energy of the $DN$ system. The energy-dependent correction term $\Delta v(W)$ is introduced to compensate the deviation from the local potential approximation. This complex and energy-dependent potential reproduces the scattering amplitude $t_{11}$ when the Schr\"odinger equation with this potential is self-consistently solved. The strength of the potential $v_{DN}(r;W)$ at $r=0$ is shown in Fig.~\ref{fig:potential}. One finds that the real part (imaginary part) is larger (smaller) than that of the $\bar{K}N$ potential~\cite{Hyodo:2007jq}, which demonstrate the differences of the interaction kernel discussed in the previous section.

\section{The Fixed Center Approximation for the $DNN$ system}\label{sec:FCA}

The fixed center approximation (FCA) to the Faddeev equations has been used with success in several problems. It is advantageous that the two-body absorption process of the three-body system can be calculated as discussed in Sec.~\ref{subsec:absorption}. One assumes that a pair of particles remains relatively unaffected by the interaction of the third particle with this pair. This usually happens when the third particle is lighter than the constituents of the pair, and also if the cluster is tightly bound. The method has been used with success in the study of $K^-$ scattering with the deuteron in Refs.~\cite{Chand:1962ec,Toker:1981zh,Kamalov:2000iy,Meissner:2006gx} (see a review in Ref.~\cite{Gal:2006cw} for comparison with full Faddeev calculations). More recently it has been applied to systems of two mesons and a baryon in Ref.~\cite{Xie:2010ig}, where the $N\bar{K}K$ system is investigated. The results obtained are in good agreement with more accurate results obtained with variational calculations in Ref.~\cite{Jido:2008kp}, or the Faddeev equations in coupled channels~\cite{MartinezTorres:2008kh,MartinezTorres:2010zv}. The puzzle of the $\Delta_{5/2^{+}}(2000)$ is also addressed with this technique, assuming this resonance to be mostly built up from $\pi \rho \Delta$ in Ref.~\cite{Xie:2011uw}.  Closer dynamically to the problem under consideration is the work~\cite{Xiao:2011rc}, where the $NDK$, $\bar{K} DN$ and $ND\bar{D}$ systems are studied with this method. 

In the present case, where we want to study the $DNN$ system, we have also the precedent of the work of Refs.~\cite{melahat,Bayar:2012rk}, where the $\bar{K}NN$ system was studied within this approximation  and found to provide results in qualitative agreement with those of the variational calculations~\cite{Dote:2008in,Dote:2008hw}. The condition that the interacting particle ($D$ meson) is lighter than those of the two-body cluster (nucleon) is not fulfilled in this case. This certainly introduces larger uncertainties than in other cases studied but we still expect that one can get good results at a qualitative level. Actually, the real difficulty of the FCA occurs when one applies it to studying possible resonant three body systems above the threshold of the three particles~\cite{MartinezTorres:2010ax}. In the present case, we look for deeply bound states of the $DNN$ system and we are safer. However, in order to be more certain about the results, we have also performed calculations using a variational method. The differences found in the two approaches can give us an idea of the uncertainties, and the features shared by the two approaches can be considered more reliable. 

\subsection{The formalism for the FCA in the $DNN$ system} 

In the FCA to the Faddeev equations for the $DNN$ three body system, one takes the  $NN$ as a cluster and $D$ scatters from  that cluster. We consider the $DNN$ system with total isospin $I_{\text{tot}}=1/2$ and with the total spin-parity $J^{P}=0^{-}$ and $J^{P}=1^{-}$. In this approach, all the two-body pairs are in $s$ wave.

First we make the evaluation for the case of $J^{P}=0^{-}$, which corresponds to the spin (isospin) of the $NN$ pair as $S_{NN}=0$ $(I_{NN}=1)$. To have total isospin $I_{\text{tot}}=1/2$, the dominant component of the $DN$ system is $I=0$, where the $\Lambda(2595)$ resonance appears. 

\begin{figure}[tbp]
\centering
\includegraphics[width=0.5\textwidth]{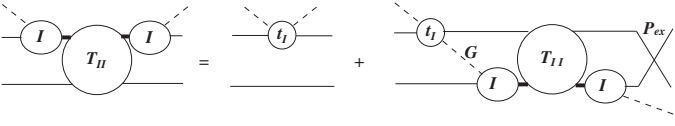}
\caption{\label{fig:DNN-T}
    Diagrammatic illustration of the three-body equation~\eqref{Teq}.}
\end{figure}

The $T$ matrix for the three-body $DNN$ scattering is labeled by the $DN$ isospins in the entrance channel $I$ and the exit channel $I'$, $T_{I,I'}$.
We denote the two-body ($s$-wave) $DN$ scattering amplitudes by $t^{(0)}$ for $I=0$ and $t^{(1)}$ for $I=1$. Then the $T$ matrix satisfies
\begin{eqnarray}
&& T_{I,I'} = t^{(I)}\delta_{I,I'} + t^{(I)} G_{I,I''}G_0  T_{I'',I'}P_{ex}
\label{Teq} ,
\end{eqnarray}
which is diagrammatically represented in Fig.~\ref{fig:DNN-T}. In Eq.~\eqref{Teq}, $G_0$ is the meson exchange propagator~\cite{multirho,melahat}
\begin{equation}
G_0=\int\frac{d^3q}{(2\pi)^3}F_{NN}(q)\frac{1}{{q^0}^2-\vec{q}\,^2-m_{D}^2+i\epsilon} ,
\label{Eq:gzero}
\end{equation}
where $F_{NN}(q)$ is the form factor, representing momentum distribution of the $NN$ system. $P_{ex}$ is the isospin exchange factor, which depends on the total isospin of the nucleon, $I_{NN}$, in the final state, $P_{ex}= (-1)^{I_{NN}+1} = 1$ for $J=0$, and $=-1$ for $J=1$.

Here we concentrate on the isospin factors in the $DNN$ scattering amplitudes. We define the isospin doublets, $N=(p,n)$, $D=(D^+, -D^0)$ and consider the $DNN$ states with the total isospin $I_{\text{tot}}=1/2$. 
There are two independent states with the total spin $J=0$ and $J=1$, which can be decomposed into the $DN$ isospin eigenstates, as
\begin{eqnarray*}
&&  
|D(N_1N_2)_{I_{NN}=1}\rangle_{J=0} =  \frac{\sqrt{3}}{2} |(DN_1)_0 N_2\rangle +\frac{1}{2} |(DN_1)_1 N_2\rangle , \\
&& 
|D(N_1N_2)_{I_{NN}=0}\rangle_{J=1} = - \frac{1}{2} |(DN_1)_0 N_2\rangle +\frac{\sqrt{3}}{2} |(DN_1)_1 N_2\rangle .
\end{eqnarray*}
The $D$ exchange matrix is given in terms of the isospin recombination factors.
\begin{eqnarray*}
&& |(DN_1)_0 N_2\rangle = \frac{1}{2} |(DN_2)_0 N_1\rangle +\frac{\sqrt{3}}{2} |(DN_2)_1 N_1\rangle ,\\
&& |(DN_1)_1 N_2\rangle = \frac{\sqrt{3}}{2} |(DN_2)_0 N_1\rangle -\frac{1}{2} |(DN_2)_1 N_1\rangle .
\end{eqnarray*}
Thus the transition matrix $G$ is given by
\begin{eqnarray*}
&& G=\begin{pmatrix}
\displaystyle \frac{1}{2}&\displaystyle\frac{\sqrt{3}}{2} \\
\displaystyle\frac{\sqrt{3}}{2}& -\displaystyle\frac{1}{2} 
\end{pmatrix} .
\end{eqnarray*}

The three-body amplitude $T_{I, I'}$ is obtained by solving Eq.~(\ref{Teq}):
\begin{align*}
T  =&  \left[
1-\frac{1}{2}(t^{(0)}-t^{(1)})G_0 P_{ex}  -t^{(0)}t^{(1)}G_0^2
\right]^{-1} \\
&\times 
\begin{pmatrix}
t^{(0)}+\frac{1}{2} t^{(1)} G_0 t^{(0)} P_{ex} & \frac{\sqrt{3}}{2} t^{(0)}t^{(1)} G_0 P_{ex} \\
\frac{\sqrt{3}}{2} t^{(0)} t^{(1)} G_0 P_{ex}& t^{(1)} -\frac{1}{2} t^{(0)} G_0 t^{(1)} P_{ex}
\end{pmatrix} .
\end{align*}
In calculating the $T$ matrix for the scatterings in the $J=0$ and $J=1$ channels, we take the linear combinations,
with a factor 2 for the choice of the first nucleon, as
\begin{eqnarray*}
&& T(J=0)= 2 
\begin{pmatrix}
\frac{\sqrt{3}}{2} & \frac{1}{2} 
\end{pmatrix} 
\begin{pmatrix}
T_{00} & T_{01}\\
T_{10} & T_{11}
\end{pmatrix} 
\begin{pmatrix}
\frac{\sqrt{3}}{2} \\
\frac{1}{2} 
\end{pmatrix} , \\
&& T(J=1)= 2 
\begin{pmatrix}
- \frac{1}{2} & \frac{\sqrt{3}}{2} 
\end{pmatrix} 
\begin{pmatrix}
T_{00} & T_{01}\\
T_{10} & T_{11}
\end{pmatrix} 
\begin{pmatrix}
-\frac{1}{2} \\
\frac{\sqrt{3}}{2} 
\end{pmatrix} .
\end{eqnarray*}
Substituting the $T$ matrix and replacing $P_{ex}$ by $+1$ for $J=0, I_{NN}=1$ scattering and $-1$ for $J=1, I_{NN}=0$, we obtain
\begin{align}
T(J=0) = &\left(\frac{3}{2} t^{(0)}+ \frac{1}{2} t^{(1)} + 2t^{(0)} t^{(1)} G_0\right) \nonumber \\
 &\times
 \left[1-\frac{1}{2}(t^{(0)}-t^{(1)})G_0  - t^{(0)}t^{(1)}G_0^2\right]^{-1} , \label{eq:TJ0}\\
 T(J=1) = &\left(\frac{1}{2} t^{(0)}+ \frac{3}{2} t^{(1)} + 2t^{(0)} t^{(1)} G_0\right) \nonumber \\
 &\times
 \left[1+\frac{1}{2}(t^{(0)}-t^{(1)})G_0   - t^{(0)}t^{(1)}G_0^2\right]^{-1} .\label{eq:TJ1}
\end{align}
These results coincide with those derived in the charge basis~\cite{Oset:2012gi,Bayar:2012rk} (see Appendix~\ref{sec:chargebasis}).

\begin{figure}[tbp]
\centering
\includegraphics[width=0.4\textwidth]{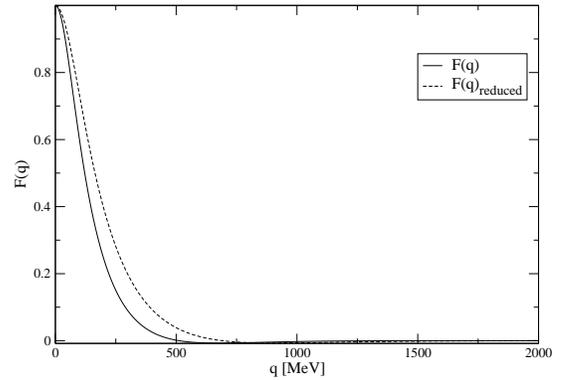}
\caption{Form factor of the deuteron, and the one corresponding to an $NN$ system with a reduced 
radius from Ref. \cite{Dote:2008hw}.}\label{fig:fff}
\end{figure}

We can see that Eq. (\ref{Eq:gzero}) contains the folding of the $D$ intermediate propagator with the form factor of the $NN$ system. The variable $q^0$ in Eq.~(\ref{Eq:gzero}) is the energy carried by the $D$, which is given by 
\begin{equation}
q^0=\frac{s+m_{D}^2-(2M_N)^2}{2\sqrt{s}} ,
\nonumber
\end{equation}
with $\sqrt{s}$ for the rest energy of the $DNN$ system. Eq. (\ref{Eq:gzero}) requires the $NN$ form factor. For $I_{NN}=0$ one could take the deuteron form factor, but the attraction of the $D$ on the nucleons will make the $NN$ system more compact, like in the case of the $\bar{K} NN$ system. Yet, there are limits on how much one can contract this system because of the strong $NN$ repulsion at short distances. In order to estimate the $NN$ size one can rely upon the results of Ref.~\cite{Dote:2008hw} in the study of the $\bar{K} NN$ system, where the $NN$ repulsion at short distance was explicitly taken into account. In practical terms we use the same expression for the form factor as for the deuteron~\cite{Machleidt:2001dd}
\begin{align}
& F(q)= \int^{\infty}_{0} d^{3}p ~\sum_{j=1}^{11} \frac{C_{j}}{\vec{p}^{2}+m_{j}^{2}}\sum_{i=1}^{11}
\frac{C_{i}}{(\vec{p}-\vec{q})^{2}+m_{i}^{2}} \label{111} ,
\end{align} 
but with the parameters $m_i$ rescaled such as to give an average separation of the nucleons of $R_{NN}\simeq 2$~fm~\cite{Dote:2008hw}. They are shown in Fig. \ref{fig:fff}. The validity of this $NN$ form factor will be examined by the result of the variational calculation, where the average distance of the $NN$ pair in the $DNN$ system will be optimized in the three-body dynamics.

We need the argument $s_1$ of the $DN$ amplitude, $t(\sqrt{s_{1}})$. To evaluate it we adopt a common procedure of dividing the binding energy into the three particles proportionally to their masses. The energy of the nucleon and the $D$ meson are given by
\begin{equation}
E_N=M_N \frac{\sqrt{s}}{2 M_N +m_D} ~,~E_D=m_D \frac{\sqrt{s}}{2 M_N +m_D},
\nonumber
\end{equation} 
so the total energy of the two-body system can be calculated as
\begin{equation}
s_1=(p_D+p_{N_1})^2=s~ \Big(\frac{M_N+m_D}{2 M_N+m_D}\Big )^2-\vec{p}_{N_2}^{~2}.
\label{Eq:arg11}
\end{equation} 
The approximate value of $\vec{p}_{N_2}^2$ can be obtained by assuming 
\begin{equation}
\frac{\vec{p}_{N_2}^{~2}}{2M_N}\simeq B_{N_2};~~B_{N_2}=M_N-M_N \frac{\sqrt{s}}{2M_N+m_D}
\label{Eq:arg12} ,
\end{equation} 
which provides a rough estimate for bound systems with the strong interaction.
  
\subsection{Evaluation of the $D (N N)$  Absorption}\label{subsec:absorption}

\begin{figure}[tbp]
\centering
\includegraphics[width=0.5\textwidth]{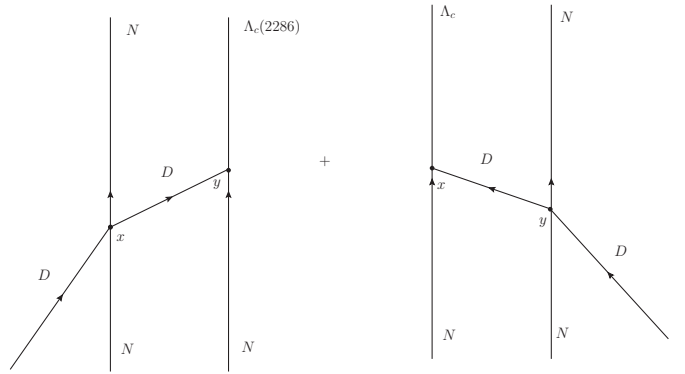}
\caption{Diagrammatic representation of the $D (N N)$  absorption.}\label{FCAfignew}
\end{figure}

As we shall see in Sec.~\ref{sec:results}, we obtain a $DNN$ bound system with a very small width. This is related to the small width of the $\Lambda_c (2598)$ state which is generated in $DN$ interaction in $I=0$. Yet, this calculation only takes into account the decay channel $DN \rightarrow \pi \Sigma_c$ for which there is little phase space and $DN \rightarrow \pi \Lambda_c$ channel which comes from the subdominant $DN$ $I=1$ component in the $DNN$ system. Now we allow the $D$ to be absorbed by two nucleons, in analogy to the $\bar{K} NN \rightarrow \Lambda N$ considered in Refs.~\cite{angelsnuc,Sekihara:2009yk}. Here the channel will be $DNN \rightarrow N \Lambda_c$ whose absorption process is shown diagrammatically in Fig. \ref{FCAfignew} (other mechanisms and decay channels will be discussed in the end of this section). We  calculate only the first diagram in Fig. \ref{FCAfignew}. The second one gives an identical contribution and they sum incoherently: there is no interference since the $N\Lambda_c$ and $\Lambda_c N$ are orthogonal states. Hence, the total width will be twice the one obtained from just one diagram.  

The $S$-matrix for the diagram is given by
\begin{eqnarray}
S&=& \int d^{4}x \int d^{4}y (-i) t_{DN\rightarrow DN} \nonumber\\
&&\times\frac{1}{\sqrt{2\omega_D}} 
\varphi_{D}(\vec x) e^{-i\omega_D x^0} e^{i E'_{N_1}x^0}e^{-i E_{N_1}x^0}
\varphi^*_{N'_1}(\vec x)\varphi_{N_1}(\vec x)\nonumber\\&&
\times\int \frac{d^{4}q}{(2\pi)^{4}}e^{-i q (y-x)} \frac{i}{q^2-m_D^2+i \epsilon} \nonumber\\
&&\times V_y \vec \sigma \vec q ~e^{i E_{\Lambda_c} y^0} e^{-i E_{N_2}y^0} \varphi^*_{\Lambda_c}(\vec y) \varphi_{N_2}(\vec y),
\nonumber
\end{eqnarray}
where $V_y$ is the Yukawa vertex. We take the same coupling as $K^- p \rightarrow \Lambda$ since in the $D$ and $\Lambda_c$ the $c$ quark plays the role of the $s$ quark in the ${\bar K}$ and $\Lambda$. In Ref.~\cite{Oset:2000eg}, the $V_y$
is given as 
\begin{align}
 V_y=-\frac{1}{\sqrt{3}}\frac{3F+D}{2 f} ,
\nonumber
\end{align} 
with $D=0.795$, $F=0.465$ \cite{Borasoy:1998pe}. We perform the $x^0$, $y^0$ integrations, and make a change of  the spatial variables as
\begin{align}
 \vec x=\vec R -\frac{\vec r}{2}, \quad
 \vec y=\vec R +\frac{\vec r}{2} . 
 \nonumber
\end{align}
Then we can write
\begin{align}
 \varphi_{N_1}(\vec x) \varphi_{N_2}(\vec y)=\frac{1}{\sqrt{V}} e^{i \vec P .\vec R} \varphi(\vec r),
\nonumber 
\end{align}
where $\varphi(\vec{r})$ is the wave function of the $NN$ system. $N'_1$ and $\Lambda_c$ will be outgoing plane waves. Let us also assume that the $D$ is a plane wave with a certain momentum. The final formula that we shall use is independent of this momentum, as we shall see. Thus,
\begin{align}
 \varphi_{D}(\vec x) =\frac{1}{\sqrt{V}} e^{i \vec p_D .\vec x} ,
 \nonumber
\end{align}
and then, using these new functions, the $S$-matrix is written as follows
 \begin{eqnarray}
S&=&\frac{1}{V^2} \int \frac{d^{3}q}{(2\pi)^{3}} \frac{1}{\sqrt{2\omega_D}} t_{DN\rightarrow DN}  
\frac{1}{q^2-m_D^2+i \epsilon}~\nonumber \\
&&\times V_y \vec \sigma \vec q~ \tilde \varphi(\vec q- \vec p_{\Lambda_c}+\frac{\vec P}{2})(2\pi)^{4}
\delta^4(p_{i}-p_{f}) \label{ss11}\\
&\equiv& -iT \frac{1}{\sqrt{2\omega_D}} \frac{1}{V^2} (2\pi)^{4}\delta^4(p_{i}-p_{f}), \nonumber
\end{eqnarray}
where $ \tilde \varphi(\vec q)$ is the Fourier transform of the wave function $\varphi(\vec r)$ normalized to 1, and $p_i$ and $p_f$ are the initial and final momentum, respectively. The $NN$ wave function in momentum space is defined as
\begin{equation}
\tilde{\varphi}(\vec{q}) = \int d^3q e^{i \vec{q~} \vec x} \varphi(\vec x), \nonumber
\end{equation}
and has a maximum value for $\vec{q}=0$. If we take the $NN$ system at rest, $\vec{P}=0$, the wave function $\tilde{\varphi}(\vec{q})$ in Eq. (\ref{ss11}) will peak at $\vec{q} - \vec{p}_{\Lambda_c} = 0$. This allows us to approximate the $D$ propagator in Eq. (\ref{ss11}) as
\begin{align}
\frac{1}{q^2-m_D^2}\rightarrow \frac{1}{(q^{0})^{2}-\vec p^{~2}_{\Lambda_c}-m_D^2},
\end{align}
where $q^0=E_{\Lambda_c}-E_{N_2}$ and $p_{\Lambda_c}\approx \lambda^{1/2}(M_{NND}^2,M_{N}^2,M_{\Lambda_c}^2)/2 M_{NND} $. We do not need to specify the $t_{DN \rightarrow DN}$ amplitude since it will be accounted for at the end of the formalism.

\begin{figure}[tbp]
\centering
\includegraphics[width=0.5\textwidth]{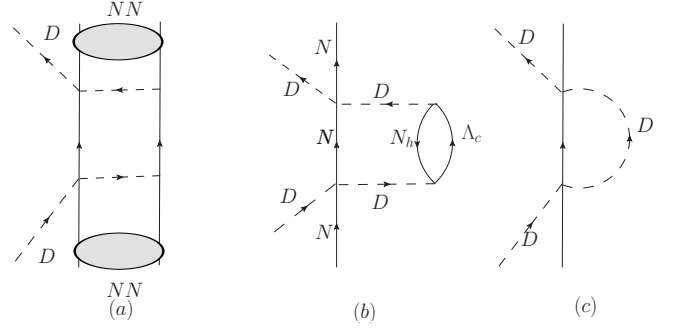}
\caption{ $D (N N)$  absorption.}\label{abs1}
\end{figure}

Defining of $\vec q - \vec p_{\Lambda_c} \equiv \vec q~'$, the square of the total matrix element is obtained as follow:
\begin{eqnarray}
 |T|^2 &=&V_y^2 \vec p^{~2}_{\Lambda_c} 
 \left(\frac{1}{(q^{0})^{2}-p_{\Lambda_c}^2-m_D^2}\right)^2 \nonumber \\
&&\times 
\left|\frac{1}{2 \pi^2}\int q'^2 dq'\tilde \varphi(\vec q~')  t_{DN,DN}(\sqrt{s'}) \right|^2\label{Eq.Tsqeski1},
 \end{eqnarray}
With this $T$ matrix we evaluate the cross section for the process of Fig. \ref{FCAfignew} (left) and we obtain
\begin{eqnarray}
 \sigma_{\text{abs}} = \frac{1}{2 \pi} \frac{ M_{NN}  M_{\Lambda_c} M_N }{M_{NND}^2 }  \frac{p_{\Lambda_c}}{p_D}|T|^2.
 \nonumber
\end{eqnarray}
It is interesting to relate this cross section to the imaginary part of the forward $D(NN) \rightarrow D(NN)$
amplitude from the diagram of Fig. \ref{abs1} using the optical theorem. We find 
\begin{eqnarray}
\text{Im }T_{D(NN)}=-\frac{p_D \sqrt{s}}{M_{NN}} \sigma_{\text{abs}} = -\frac{1}{2 \pi} \frac{M_{\Lambda_c} M_N }{M_{NND}} p_{\Lambda_c} |T|^2. \nonumber
\end{eqnarray}

The next step is to convert the absorption diagram of the Fig. \ref{abs1} (a) into a ``many body'' diagram of Fig. \ref{abs1} (b) where the nucleon where the $D$ is absorbed, the only occupied state of the ``many body'' system, is converted into a hole state in the many body terminology \cite{walecka}. Once this is done one observes that if we remove the amplitude $t_{DN}$ in the expression of $T$, the expression that we obtain for $\text{Im }T_{D(NN)}$ corresponds to the evaluation of the imaginary part of the two-body loop function $g$ of a nucleon and a $D$ meson [Fig. \ref{abs1} (c)] but with a $D$ selfenergy insertion accounting for the ($\Lambda_c N_h$) excitation of the $D$ meson. We call this $\delta \tilde{g} $. The Feynman rules to evaluate $\text{Im }\delta \tilde{g}$ and $\text{Im }T_{D(NN)}$ are identical, except that $t_{DN,DN}$ is removed in the evaluation of $\text{Im }\delta \tilde{g} $. Hence we obtain
\begin{eqnarray}
 i \text{Im }\delta \tilde{g}= -i \frac{1}{2 \pi} \frac{M_{\Lambda_c} M_N }{M_{NND}} p_{\Lambda_c} |\tilde{T}|^2.
\nonumber
\end{eqnarray} 
with  $|\tilde{T}|^2$ is given by Eq. (\ref{Eq.Tsqeski1}) removing $t_{DN,DN}$. This simplifies the expression since 
\begin{align}
\frac{1}{2 \pi^2}\int q'^2 dq'\tilde \varphi(\vec q~')
=& \lim_{r \rightarrow 0}\int\dfrac{d^3q'}{(2\pi)^3} e^{i\vec q~'\vec r } \tilde \varphi(\vec q~') \\
=&\varphi(r=0)  .
\nonumber
\end{align} 
Thus $|\tilde{T}|^2$ is given by  
\begin{eqnarray}
 |\tilde{T}|^2 &=&V_y^2 \vec p^{~2}_{\Lambda_c}\frac{1}{[(q^{0})^{2}-p_{\Lambda_c}^2-m_D^2]^2} |\varphi(0)|^{2} \nonumber .
\end{eqnarray}
Finally $\vec p^{~2}_{\Lambda_c}$ accompanying $V_y^2$ in the former expression requires a small correction. The factor comes from the non relativistic $ \vec \sigma  \vec q  $ form of the $ DN \Lambda_c $ vertex. If we take instead the relativistic Yukawa vertex of the type $ \gamma^\mu \gamma^5 $, then we find the easy prescription to account for the relativistic correction,
\begin{eqnarray}
 V_y^2 \vec p^{~2}_{\Lambda_c} & \rightarrow &V_y^2 \frac{1}{4 m_{\Lambda_c}^2} (M_N+M_{\Lambda_{c}})^2 \vec p^{~2}_{\Lambda_c}.
 \nonumber
\end{eqnarray}
 
The next step is to reevaluate the $ t_{DN,DN} $ amplitude used as input in the fixed center formulas. As we mentioned, they were obtained using the method of Ref. \cite{mizuangels} with several coupled channels and the formula~\eqref{eq:Tamp}. We redo the evaluation by replacing the loop function in the $ DN $ channel as
\begin{equation}
g_{DN} \rightarrow g_{DN}+i ~\text{Im }\delta \tilde{g}
\label{Eq:bethesalpeter1}
\end{equation}
to take into account the $D$ absorption by two nucleons or, analogously, the $ \Lambda_c N_h $ excitation of the $ D $ meson. When doing this, the $ DN $ amplitude becomes complex below the  $ DN $ threshold and the narrow  $ \Lambda_c (2598) $ resonance acquires now a moderate width due to the $ D $ absorption with a second nucleon. The second process of Fig. \ref{FCAfignew}  (right) is accounted for when we consider the three-body amplitude $T$ in the FCA formula with the first $ D $ scattering with the second nucleon. 

\begin{figure}
\centering
\includegraphics[width=0.45\textwidth]{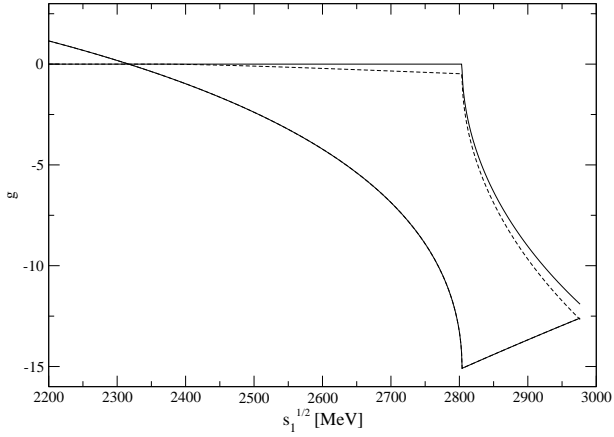}
\caption{The meson-baryon loop function $g_{DN}$ in the $DN$ channel (solid line) and with the effect of the two-body absorption $i\text{Im }\delta\tilde{g}$ added (dashed line). }\label{fig:gfunc}%
\end{figure}%

For the estimation of the width we take the wave function
\begin{align*}
\tilde{\varphi} (r)=&a e^{-\alpha r}, \quad a = \frac{1}{2} 
\left(\frac{\alpha^3}{2\pi}\right)^{\frac{1}{2}}, \\
\tilde{\varphi} (q) =& \frac{4 \pi a \alpha}{(\frac{1}{4} \alpha^2 - q^2)^2 + q^2\alpha^2},
\end{align*}
with $\alpha \simeq 1.7 \text{fm}^{-1}$, which corresponds to an $NN$ object of relative distance $2\,\text{fm}$.

\begin{figure}
\centering
\includegraphics[width=0.45\textwidth]{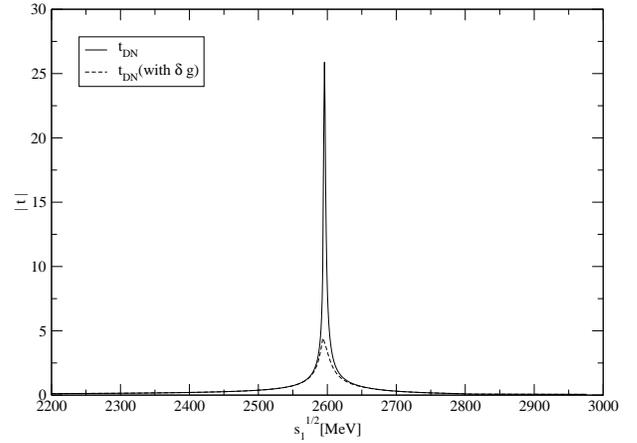}
\caption{Modulus of the two-body amplitude $DN \rightarrow DN$ (solid line) and with the effect of the two-body absorption $i\text{Im }\delta\tilde{g}$ added (dashed line).}\label{fig:tdn}
\end{figure}

Let us numerically investigate the effect of the absorption using the model described in Sec.~\ref{subsec:amplitude}. In Fig. \ref{fig:gfunc}, we show the meson-baryon loop function $g_{DN}$ in the $DN$ channel together with the two-body absorption contribution to the imaginary part, $i\text{Im }\delta\tilde{g}$. We can see that the imaginary part of the total $g$ function is no longer zero below the $ DN $ threshold due to $ D $ absorption. In Fig. \ref{fig:tdn}, we show the modulus of the two-body amplitude $|t|$ for the $ DN $ channel for $I=0$ using $g_{DN}$ and $g_{DN}+i\text{Im }\delta\tilde{g}$ of Eq.~(\ref{Eq:bethesalpeter1}). As we can see, the inclusion of the absorption mechanism induces an increase in the width of the peak of $\Lambda_{c}(2595)$ in $|t|$  which will have repercussion in the width of the $DNN$ system.

For a narrow resonance, we can approximate the amplitude around the resonance energy by a Breit-Wigner form 
\begin{equation}
t(\sqrt{s_{1}})\simeq \dfrac{g^{2}}{\sqrt{s_{1}}-M_{R}+i\frac{\Gamma}{2}} .
\nonumber
\end{equation}
This leads to the expression of the coupling of the resonance to the $DN$ scattering state as
\begin{equation}
g^{2}=\frac{1}{2} \Gamma |t(M_{R})|.
\nonumber
\end{equation}
Inspection of Fig.~\ref{fig:tdn}, together with the values of $ \Gamma $(no absorption)$ = $ 3 MeV and $ \Gamma $(absorption)$ = $ 15 MeV, show that the value of the coupling $ g^{2} $ barely changes from the introduction of $i\text{Im }\delta\tilde{g}$, but of course the resonance has become wider. Indeed  $ g^{2} $(no absorption)/ $ g^{2} $(absorption)$ \simeq 6/5$.

The absorption diagram that we have considered is not the only one, but it is the most relevant. On the same footing we should consider the diagrams where the Yukawa coupling produces $\Sigma_c$ or even $\Sigma^{*}_c(2520)$. The analogy with the kaons made before, and the values of these couplings that can be seen in Ref.~\cite{Borasoy:1998pe}, together with the dynamical factor $p^{3}$ of the cross section, make the contribution of these terms of the order of 5\% of the $\Lambda_c$ production and we neglect them. Analogously we can also have $DN \to \pi (\eta) \Lambda_c (\Sigma_c)$ in the first hadron line of the absorption diagram and exchange a pion or an eta. These diagrams are further suppressed because they require the exchange of a heavy vector in the $DN \to \pi(\eta) Y_c$ amplitude in the extension of the hidden gauge approach that we use. They are penalized by the factor $\kappa_c ^2$, which, even considering that the pion (eta) propagators have bigger strength than the $D$ one, renders these diagrams at the level of 10\%.

\section{Variational calculation of the $DNN$ system}\label{sec:variational}

Here we calculate the energy of the $DNN$ system with a variational approach formulated for $\bar{K}NN$ system in Refs.~\cite{Dote:2008in,Dote:2008hw}. As in the case of the FCA, we consider the $DNN$ system with total isospin $I=1/2$ and the total spin-parity either $J^{P}=0^{-}$ or $J^{P}=1^{-}$. The trial wave function for the $J^{P}=0^{-}$ state is prepared with two components:
\begin{equation}
    \ket{\Psi^{J=0}} 
    = (\mathcal{N}^{0})^{-1}[\ket{\Phi_{+}^{0}}
    +C^{0}\ket{\Phi_{-}^{0}}] ,
    \nonumber
\end{equation}
where $\mathcal{N}^{0}$ is a normalization constant and $C^{0}$ is a mixing coefficient. In the main component $\ket{\Phi_{+}^{0}}$, two nucleons are combined into spin $S_{NN}=0$ and isospin $I_{NN}=1$ so all the two-body subsystems can be in $s$ wave. We also allow a mixture of the $\ket{\Phi_{-}^{0}}$ component where both spin and isospin are set to be zero, so the orbital angular momentum between two nucleons is odd. The $J^{P}=1^{-}$ state is studied in a similar way as
\begin{equation}
    \ket{\Psi^{J=1}} 
    = (\mathcal{N}^{1})^{-1}[\ket{\Phi_{+}^{1}}
    +C^{1}\ket{\Phi_{-}^{1}}] ,
    \nonumber
\end{equation}
where $\ket{\Phi_{+}^{1}}$ ($\ket{\Phi_{-}^{1}}$) denotes $S_{NN}=1$ and $I_{NN}=0$ ($S_{NN}=1$ and $I_{NN}=1$) component. Note that only the main component of $\ket{\Phi_{+}^{J=0,1}}$ is taken into account in the FCA calculation. The wave functions are expanded in terms of gaussians in coordinate space, and we minimize the total energy of the system with the Hamiltonian given below. Detailed explanation of the variational method can be found in Ref.~\cite{Dote:2008hw}.

We consider the following Hamiltonian in this study:
\begin{equation}
    \hat{H}
    = \hat{T}+\hat{V}_{NN}+\re\hat{V}_{DN}
    -\hat{T}_{\text{c.m.}} ,
    \label{eq:Hamiltonian}
\end{equation}
where $\hat{T}$ is the total kinetic energy, $\hat{V}_{DN}$ is the $DN$ potential term which is the sum of the contributions from two nucleons, and $\hat{T}_{\text{c.m.}}$ is the energy of the center-of-mass motion. For the $NN$ potential $\hat{V}_{NN}$, we use three models: HN1R which is constructed from Hasegawa-Nagata No.1 potential~\cite{PTP45.1786}, the Minnesota force~\cite{Thompson:1977zz}, and the gaussian-fitted  version of the Argonne v18 potential~\cite{Wiringa:1994wb}. The characteristic features of these $NN$ potentials are summarized in Appendix~\ref{sec:NNint}. For later convenience, we define the following matrix elements
\begin{align}
    E_{\text{kin}}
    =&
    \bra{\Psi}\hat{T}-\hat{T}_{\text{c.m.}}\ket{\Psi} ,
    \nonumber \\
    V(NN)
    =&
    \bra{\Psi}\hat{V}_{NN}\ket{\Psi} ,
    \nonumber \\
    V(DN)
    =&
    \bra{\Psi}\re \hat{V}_{DN}\ket{\Psi} ,
    \nonumber \\
    T_{\text{nuc}}
    =&
    \bra{\Psi}\hat{T}_{N}-\hat{T}_{\text{c.m.},N}\ket{\Psi} ,
    \nonumber \\
    E_{NN}
    =&
    T_{\text{nuc}}+V(NN) ,
    \nonumber 
\end{align}
where $\hat{T}_{N}$ and $\hat{T}_{\text{c.m.},N}$ are nucleonic parts of the kinetic term and center-of-mass energy, respectively. 

We take the real part of the $DN$ potential for the energy variation, and the imaginary part will be used to estimate the mesonic decay width. The energy dependence of the interaction was treated self-consistently in the study of $\bar{K}NN$ system~\cite{Dote:2008hw}. While the $\bar{K}N$ amplitude is well calibrated by experimental data such as total cross sections and $\pi\Sigma$ mass distributions, the $DN$ amplitude is only constrained by the mass of the quasi-bound state $\Lambda_{c}(2595)=\Lambda_{c}^{*}$. In addition, the self-consistent treatment requires some assumption on the energy fraction of the $DN$ pair in the three-body system, which cannot be determined unambiguously. In this study, therefore, we refrain from the self-consistent treatment of the energy of the $DN$ subsystem and set the strength of the potential at the energy of $\Lambda_{c}^{*}$ resonance:
\begin{align}
    \re v_{DN}(r=0;W=M_{\Lambda_{c}^{*}})
    =&
    \begin{cases}
    -1336 \text{ MeV} & (I=0) \\
    -343 \text{ MeV} & (I=1)
    \end{cases} ,
    \label{eq:strength}
\end{align}
with $M_{\Lambda_{c}^{*}}=2597.1$ MeV. In this case, the $M_{\Lambda_{c}^{*}}$ in $I=0$ channel is correctly reproduced, while the $I=1$ resonance disappears, because the strength of the $DN$ potential~\eqref{eq:DNpotential} reduces at the lower energy region as seen in Fig.~\ref{fig:potential}.

It is useful to introduce one- and two-body densities in order to extract the spatial structure of the $DNN$ bound state. We first define the one-body densities as 
\begin{align}
    \rho_{N}(r)
    =&
    \bra{\Psi}
    \sum_{i=1,2}\delta^{3}(|\bm{r}_{i}-\bm{R}_{G}|-r)
    \ket{\Psi} ,
    \nonumber \\
    \rho_{D}(r)
    =&
    \bra{\Psi}
    \delta^{3}(|\bm{r}_{D}-\bm{R}_{G}|-r)
    \ket{\Psi} ,
    \nonumber \\
    \rho_{T}(r)
    =&
    \rho_{N}(r)+\rho_{D}(r)
    \nonumber ,
\end{align}
where $\bm{R}_{G}$ is the center-of-mass coordinate of the three-body system. The one-body densities represent the probability of finding $N$, $D$, or any of them at distance $r$ from the center of mass of the system. We also define the two-body correlation densities as 
\begin{align}
    \rho_{NN}(x)
    =&
    \bra{\Psi}
    \delta^{3}(|\bm{r}_{1}-\bm{r}_{2}|-x)
    \ket{\Psi} ,
    \nonumber \\
    \rho_{DN}(x)
    =&
    \bra{\Psi}
    \sum_{i=1,2}\delta^{3}(|\bm{r}_{D}-\bm{r}_{i}|-x)
    \ket{\Psi} ,
    \nonumber 
\end{align}
which stand for the probabilities of finding $NN$ or $DN$ pair at relative distance $x$. The root-mean-square radius of particle $X$, $\sqrt{\langle r^{2}\rangle_{X}}$, and relative distance of particles $X$ and $Y$, $R_{XY}$, are given as the second moment of the one-body and two-body densities, respectively:
\begin{align}
    \langle r^{2}\rangle_{X}
    =&
    \int d^{3}\bm{r}\ \bm{r}^{2}\rho_{X}(r)   , \nonumber \\
    R_{XY}^{2}
    =&
    \int d^{3}\bm{x}\ \bm{x}^{2}\rho_{XY}(x)  .
    \nonumber 
\end{align}

In this setup, since the imaginary part of the $DN$ potential is not included, the $\Lambda_{c}^{*}$ appears as a stable bound state. Thus, in the variational approach, the $DNN$ three-body bound state can be found in the energy region below the $\Lambda_{c}^{*}N$ threshold $\sqrt{s}\sim 3536$ MeV. If the three-body (quasi-)bound state exists above the $\Lambda_{c}^{*}N$ threshold, variational calculation will find the $\Lambda_{c}^{*}N$ two-body scattering state as the ground state of the three-body system.

A three-body bound state above the $\pi\Lambda_{c}N$ threshold $\sqrt{s}\sim 3363$ MeV has a mesonic decay width. The three-body decay width can be estimated by the matrix element of the imaginary part of the $DN$ potential as
\begin{align}
    \Gamma_{\pi Y_{c}N}
    =&
    -2 \bra{\Psi}\im \hat{V}_{DN}\ket{\Psi}
    \nonumber ,
\end{align}
where $\ket{\Psi}$ is the obtained wave function of the ground state. As seen in Fig.~\ref{fig:potential}, the imaginary part of the $DN$ potential is much smaller than the real part. This may justify the perturbative treatment of the imaginary part, which ignores the dispersive effect on the energy of the $DNN$ system from the imaginary part.
 
\section{Results}\label{sec:results}

\subsection{Quasi-bound states in the FCA approach}\label{subsec:FCAresult}

\begin{figure}
\begin{center}
\includegraphics[width=0.45\textwidth]{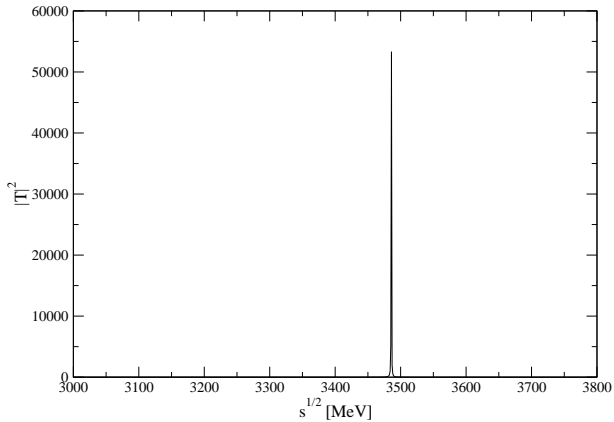}
\caption{Modulus squared of the three-body scattering amplitude for $I=1/2$ and $J=0$ with reduced size of the $NN$ radius.}
\label{fig:tmats0}
\end{center}
\end{figure}

\begin{figure}
\begin{center}
\includegraphics[width=0.45\textwidth]{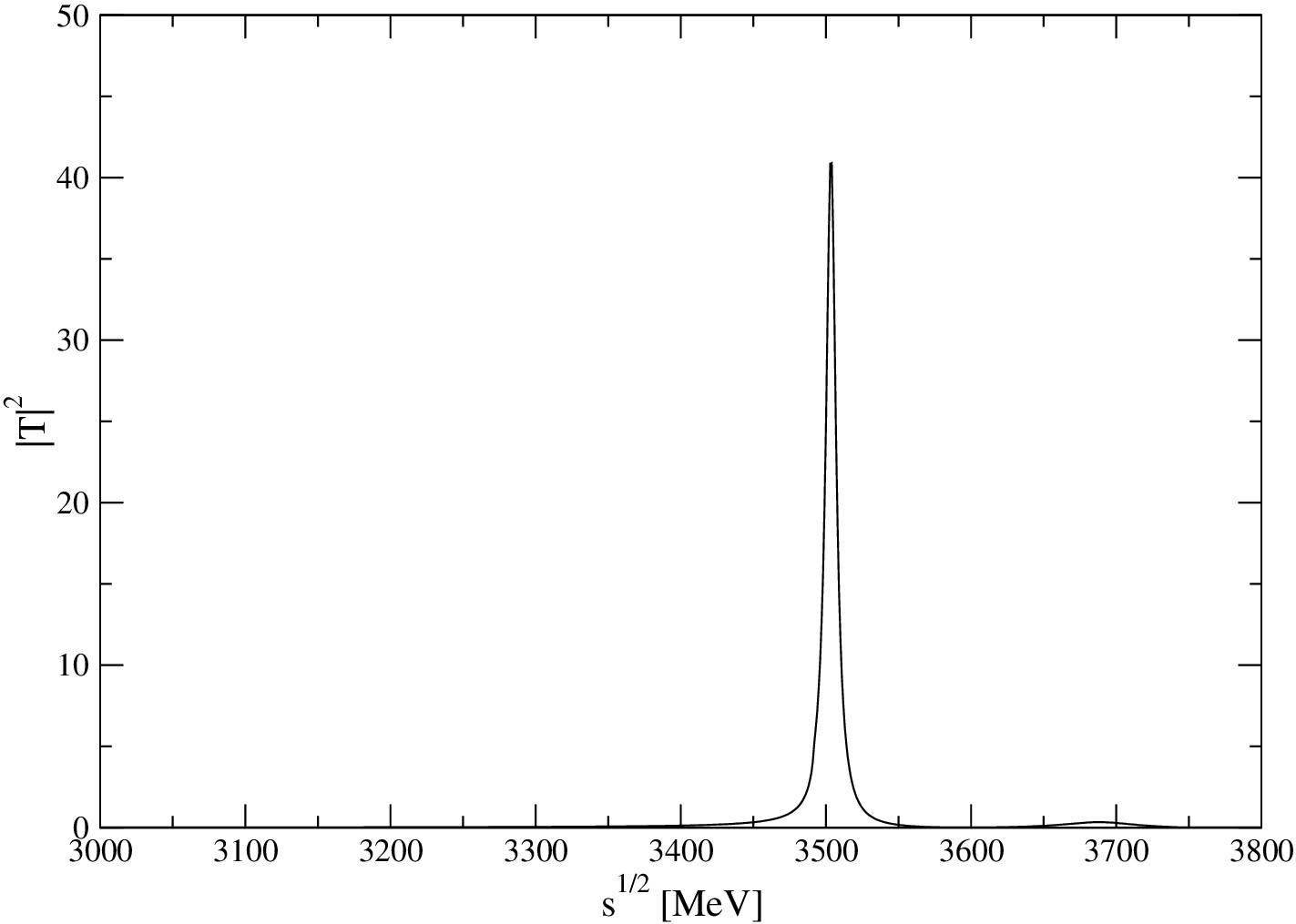}
\caption{Modulus squared of the three-body scattering amplitude for $I=1/2$ and $J=1$ with reduced size of the $NN$ radius\\$~~$.}
\label{fig:tmats1}
\end{center}
\end{figure}

\begin{figure}
\begin{center}
\includegraphics[width=0.45\textwidth]{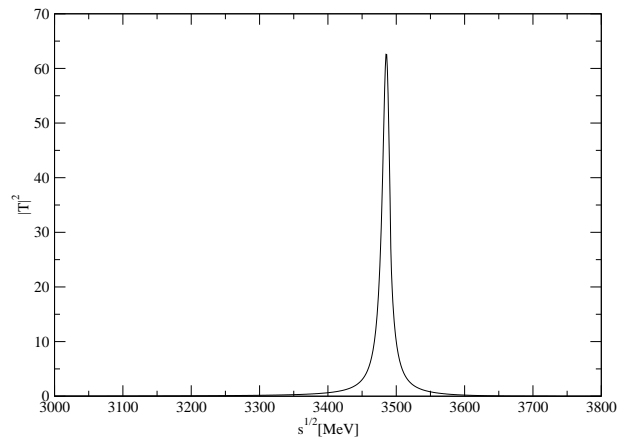}
\caption{Modulus squared of the three-body scattering amplitude for $I=1/2$ and $J=0$ (with $\delta \tilde{G}$) with reduced $NN$ radius.}
\label{fig:redtmats0}
\end{center}
\end{figure}%

\begin{figure}
\begin{center}$DN \rightarrow DN$
\includegraphics[width=0.45\textwidth]{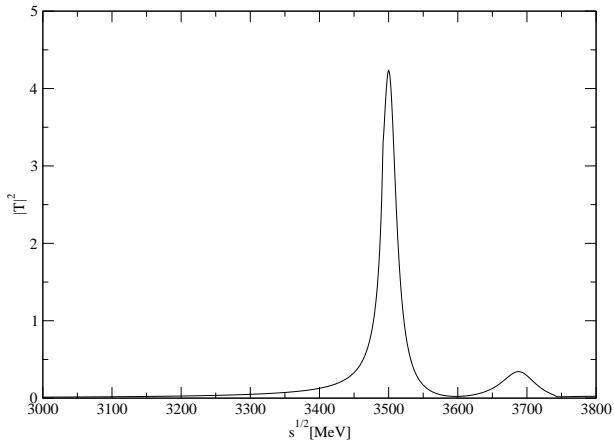}
\caption{Modulus squared of the three-body scattering amplitude for $I=1/2$ and $J=1$ (with $\delta \tilde{G}$) with reduced $NN$ radius.}
\label{fig:redtmats1}
\end{center}
\end{figure}

We first study the quasi-bound state found in the FCA calculation. In Figs.~\ref{fig:tmats0} and \ref{fig:tmats1} we show the results  for $|T|^2$ as functions of the total energy $\sqrt{s}$ assuming the $NN$ system to have reduced size. Both for $I_{NN} = 0,\, I_{NN} = 1 (J=1,\,J=0)$, we obtain a neat peak. The resonance energy for $J=0$ is about 3486 MeV and the width is extremely small. In the case of $J=1$ we have a smaller binding and the energy is about 3500 MeV, with a width of around 9 MeV. We should note that the binding is similar for both the spin channels. The position of the peak in this approximation is, in a rough estimate, given by the position of the pole of the $\Lambda_c(2595)$. This gives the value of $s_1$ and through Eqs. (\ref{Eq:arg11}), (\ref{Eq:arg12}) the value of $s$. 

However, one should note the different strength of  $ |T|^2 $ in these two cases, but a direct comparison cannot be done because the strength of the resonance amplitude at the peak is related to the width, which strongly depends on the spin. A proper comparison is better done after the $D$ absorption is included where the widths are similar.

Next we include the $\delta \tilde{g}$ to account for absorption and plot $ |T|^2 $ for the $ DNN $ system in Figs.~\ref{fig:redtmats0} and \ref{fig:redtmats1} for $J = 0$ ($I_{NN}=1$) and $J = 1$ ($I_{NN}=0$). The difference of the peak position by the absorption effect is only a few MeV (2-4 MeV) which is certainly within our uncertainties. The novelty, which is welcome, is that $ |T|^2 $ has become now wider and acquires a width of about 20-25 MeV. We are now in a position to compare the strength of these two amplitudes and we see that in the case of $J = 0$ the strength of $ |T|^2 $ at the peak is about a factor 15 larger than that for $J = 1$. This means that the state that we find at $J = 1$ should be more difficult to see, or alternatively we should see the small strength as an indication that this state is more uncertain in our approximation, as should be the smaller shoulder that one can see at higher energies for $J = 1$ in Fig.~\ref{fig:redtmats1}.

\subsection{Quasi-bound states in the variational approach}\label{subsec:variationalresult}

Now we investigate the same system in the variational approach. We first adopt HN1R potential for the nuclear force. As a result of the variational calculation, we have found that the total spin $J=1$ system ($I_{NN}=0$) is unbound with respect to the $\Lambda_{c}^{*}N$ threshold. A bound state of spin $J=0$ system ($I_{NN}=1$) is found at
\begin{equation}
    B
    \sim
    225 \text{ MeV} 
    ,
    \nonumber
\end{equation}
measured from the $DNN$ threshold ($\sim 3745$ MeV). This corresponds to the total energy of the three-body system as
\begin{equation}
    M_{B}
    \sim  3520
    \text{ MeV} .
    \nonumber
\end{equation}
We also examine the Minnesota force and Av18 potential. The results are summarized in Table~\ref{tab:energy}, together with the contributions from the individual terms in Eq.~\eqref{eq:Hamiltonian}. 

\begin{table}[tdp]
\caption{Results of the energy compositions in the variational calculation for the ground state of the $DNN$ system with total isospin $I=1/2$ (range parameter $a_{s}=0.4$ fm). Terms ``bound'' and ``unbound'' are defined with respect to the $\Lambda_{c}^{*}N$ threshold. All the numbers are given in MeV.}
\begin{center}
\begin{ruledtabular}
\begin{tabular}{lrrrr}
                      & HN1R     &          & Minnesota & Av18 \\
                      & $J=1$    & $J=0$    & $J=0$     & $J=0$  \\
\hline
                      & unbound  & bound    & bound     & bound \\
$B$                   & 208      & 225      & 251       & 209 \\
$M_{B}$               & 3537     & 3520     & 3494      & 3536 \\
$\Gamma_{\pi Y_{c}N}$ & -        & 26       & 38        & 22 \\[5pt]
$E_{\text{kin}}$      & 338      & 352      & 438       & 335 \\
$V(NN)$               & 0        & $-2$     & 19        & $-5$\\
$V(DN)$               & $-546$   & $-575$   & $-708$    & $-540$ \\
$T_{\text{nuc}}$      & 113      & 126      & 162       & 117 \\
$E_{NN}$              & 113      & 124      & 181       & 113 \\[5pt]
$P(\text{Odd})$       & 75.0 \%  & 14.4 \%  & 7.4 \%    & 18.9 \% \\
\end{tabular}
\end{ruledtabular}
\end{center}
\label{tab:energy}
\end{table}%

As seen in the Table~\ref{tab:energy}, the $DNN$ system in the $J=0$ channel is bound below the $\Lambda_{c}^{*}N$ threshold ($B\sim 209$ MeV) for all the $NN$ potentials employed.\footnote{Av18 case is almost at the $\Lambda_{c}^{*}N$ threshold, but we confirm that the wave function is localized as we will see in Sec.~\ref{subsec:structure}.} A large kinetic energy of the deeply bound system is overcome by the strong attraction of the $DN$ potential, while the $NN$ potential adds a small correction. Comparing the results with three different nuclear forces, we find that the binding energy is smaller when the $NN$ potential has a harder repulsive core (see Appendix~\ref{sec:NNint}).

In the $J=1$ channel, the ground state energy is obtained slightly above the $\Lambda_{c}^{*}N$ threshold. The fact that the $J=1$ channel is unbound is confirmed by changing the parameter $\mu$ in the trial wave function, which controls the size of the total system~\cite{Dote:2008hw}. By increasing the system size, the total energy gradually approaches the $\Lambda_{c}^{*}N$ threshold. This indicates that the lowest-energy state is indeed a two-body scattering state of the $\Lambda_{c}^{*}N$ channel. A large fraction of the odd component in this channel ($\sim 75$ \%) is realized to enhance the $I_{NN}=1$ component which has larger fraction of the $I_{DN}=0$ than the $I_{NN}=0$ component. In fact, pure $\ket{(DN)_{I=0}N}$ state can be decomposed into $I_{NN}=0$ and $I_{NN}=1$ components with the ratio 1:3. Since the $I_{NN}=1$ state is the odd state in $J=1$ ($S_{NN}=1$) channel, the $75$ \% fraction of the odd component indicates that the $DN$ pair forms the $\Lambda_{c}^{*}$. We also examine the $J=1$ channel with the Minnesota force. Although the repulsive core is soft in this case, no bound $\Lambda_{c}^{*}N$ is found. 

Using the imaginary part of the $DN$ potential, we evaluate the mesonic decay width of the quasi-bound state in the $J=0$ channel, $\Gamma_{\pi Y_{c}N}$. The results are 20-40 MeV as shown in Table~\ref{tab:energy}. This corresponds to the result of FCA without the $D$ absorption, where the width is less than 10 MeV. Note, however, that in the variational approach we have evaluated the width perturbatively, while in the FCA the evaluation is done nonperturbatively. In this sense, $\Gamma_{\pi Y_{c}N}$ obtained in the variational approach can only be regarded as an estimation of the mesonic decay width.

\subsection{Structure of the $DNN$ quasi-bound state}\label{subsec:structure}

To further investigate the structure of the $DNN$ systems, we calculate the expectation values of various distances of the obtained wave function. The results of the root-mean-square radii and the relative distances are shown in Table~\ref{tab:structure}. Except for the Av18 case where the wave function spreads due to the weaker binding, the size of the $DNN$ bound state in the $J=0$ channel is smaller than the $\bar{K}NN$ system, in which the $NN$ and $\bar{K}N$ distances are $R_{NN}\sim 2.2$ fm and $R_{\bar{K}N}\sim 1.9$ fm. It is, on the other hand, acceptable to use the reduced size of Eq.~\eqref{111} for the $NN$ distribution in the FCA calculation, given the uncertainty that arises from the choice of the $NN$ interaction. The large relative distances in the $J=1$ channel also reflect the nature of the scattering state in this channel. 

\begin{table}[tdp]
\caption{Structure of the $DNN$ ground state (range parameter $a_{s}=0.4$ fm). $\sqrt{\langle r^2 \rangle_T}$, $\sqrt{\langle r^2 \rangle_D}$ and $\sqrt{\langle r^2 \rangle_N}$ mean the root-mean-square radius of the distribution of total system, nucleons and $D$ meson, respectively. $R_{NN}$ ($R_{DN}$) is  the mean distance between two nucleons ($D$ meson and a nucleon) in the $DNN$. $R_{DN}(I)$ is the mean distance of a $DN$ component 
with isospin $I$. All the numbers are given in fm.}
\begin{center}
\begin{ruledtabular}
\begin{tabular}{lrrrr}
              & HN1R  &   & Minnesota & Av18 \\
              & $J=1$ & $J=0$ & $J=0$ & $J=0$ \\
\hline
$\sqrt{\langle r^{2}\rangle_{T}}$ 
              & 4.81  & 0.75  & 0.50  & 1.26 \\
$\sqrt{\langle r^{2}\rangle_{N}}$ 
              & 5.61  & 0.88  & 0.59  & 1.47 \\
$\sqrt{\langle r^{2}\rangle_{D}}$ 
              & 2.52  & 0.41  & 0.28  & 0.67 \\[5pt]
$R_{NN}$      & 10.04 & 1.55  & 1.03  & 2.62 \\
$R_{DN}$      & 7.11  & 1.12  & 0.76  & 1.87 \\[5pt]
$R_{DN}(I=0)$ & 4.52 & 0.83  & 0.62  & 1.28 \\
$R_{DN}(I=1)$ & 10.03 & 1.57  & 1.03  & 2.65 \\
\end{tabular}
\end{ruledtabular}
\end{center}
\label{tab:structure}
\end{table}%

\begin{figure}[tbp]
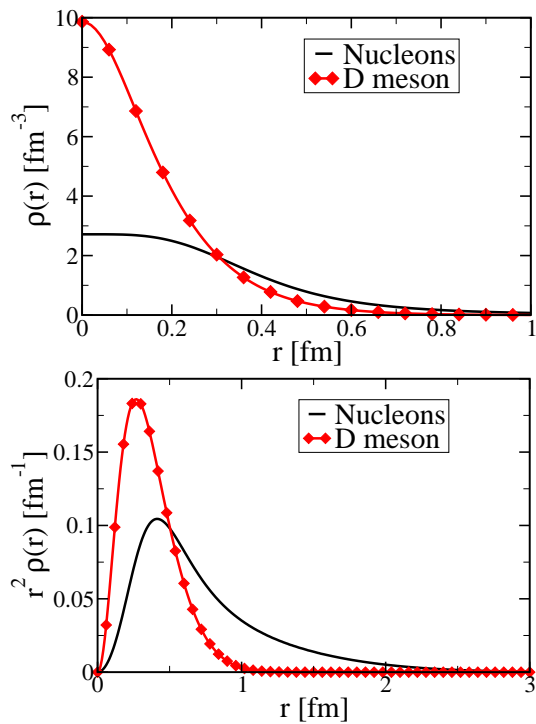

    \centering
    \includegraphics[width=7cm,clip]{Dens1DN.eps}
    \includegraphics[width=7cm,clip]{Dens1DN_r2N.eps}
    \caption{\label{fig:onebody}
    (Color online) Top: One-body densities $\rho_{N}(r)$ and $\rho_{D}(r)$ in the $J=0$ channel with HN1R potential. Bottom: the same plot of the densities multiplied by $r^{2}$.}
\end{figure}

In Fig.~\ref{fig:onebody}, we show the one-body densities of the nucleon and $D$ meson of the quasi-bound state with the HN1R potential. It is clear that the $D$ meson distributes more compactly than the nucleons. This result indicates a schematic picture where the $D$ meson sits at the center and nucleons circulates around it.

\begin{figure}[tbp]
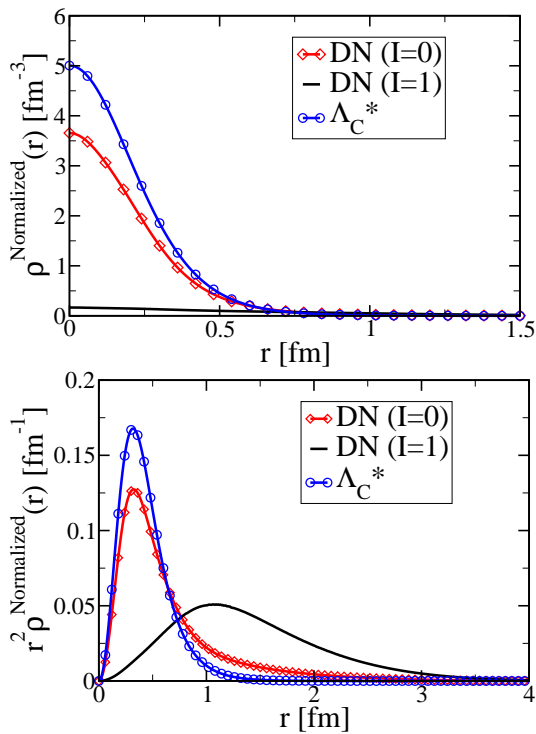

    \centering
    \includegraphics[width=7cm,clip]{Dens2DNiso_N.eps}
    \includegraphics[width=7cm,clip]{Dens2DNiso_Nr2.eps}
    \caption{\label{fig:DNtwobody}
    (Color online) Top: Normalized $DN$ two-body correlation density $\rho_{DN}(r)$ with isospin decomposition. The $I=0$ $DN$ bound state ($\Lambda_{c}^{*}$) correlation density is also shown for comparison. Bottom: the same plot of the densities multiplied by $r^{2}$.}
\end{figure}

It is instructive to look at the $DN$ correlation in more detail. In Fig.~\ref{fig:DNtwobody}, we show the $DN$ two-body correlation density as well as its isospin decomposition. It is seen that the $I=0$ component distributes more compactly than the $I=1$ component, which reflects the strength of the attraction in each channel [see Eq.~\eqref{eq:strength}]. Moreover, the $I=0$ component is similar to the distribution of the relative distance of the $DN$ two-body bound state $\rho_{\Lambda_{c}^{*}}(r)$. This indicates that the structure of the $\Lambda_{c}^{*}$ is maintained even in the three-body system. This feature has also been found in the $\bar{K}NN$ system~\cite{Yamazaki:2007cs,Dote:2008hw}.

As in the case of the $\bar{K}NN$ system, the survival of the $\Lambda_{c}^{*}$ in the three-body system opens the possibility of the ``$\Lambda_{c}^{*}$-hypernuclei'', in which the $\Lambda_{c}^{*}$ is treated as an effective degrees of freedom~\cite{Arai:2007qj,Uchino:2011jt}. In fact, this picture is more suitable in the charm sector, since the width of the $\Lambda_{c}^{*}$ is smaller than the $\Lambda^{*}$ so the effect of the imaginary part in the calculation should be smaller. Note also that the binding of the $DN$ system is as large as 200 MeV, while the binding of the $\Lambda_{c}^{*}N$ is much smaller, especially for the case of the realistic Av18 potential.

We have examined theoretical uncertainties in the construction of the potential. The range parameter of the $DN$ potential $a_{s}$ is introduced in Eq.~\eqref{eq:DNpotential} and chosen to be $0.4$ fm. When we adopt $a_{s}=0.35$ fm, the binding energy changes by a few MeV, and the size changes less than 0.1 fm. The Minnesota potential has a parameter $u$ which controls the strength of the $NN$ odd force~\cite{Thompson:1977zz}. The effect of the slight inclusion of the odd force ($u=0.95$) turns out to be very small, less than 1 MeV. We thus conclude that these uncertainties are much smaller than the dependence on the choice of the $NN$ potential. The variation of the values in Tables~\ref{tab:energy} and \ref{tab:structure} can be regarded as the theoretical uncertainties in the present calculation.

\section{Discussion}\label{sec:discussion}

\subsection{Comparison of two approaches}\label{subsec:comparison}

We have presented the results of two approaches, the Faddeev FCA calculation and the variational calculation. In the total spin $J=0$ channel, both approaches find a quasi-bound state around 3500 MeV which is below the $\Lambda_{c}^{*}N$ threshold. The assumed $NN$ distribution in the FCA turns out to be similar with that found in the variational calculation by minimizing the total energy. It is therefore reasonable to conclude that these approaches find the same quasi-bound state.

The spin $J=1$ channel, on the other hand, has differences in the two approaches. The lowest-energy state obtained in the variational calculation is a $\Lambda_{c}^{*}N$ scattering state, while a narrow peak is found in the FCA amplitude below the $\Lambda_{c}^{*}N$ threshold, although the signal strength is not so significant as the $J=0$ case. A major reason of this discrepancy may be traced back to the $DN$ interaction in the isospin $I=1$ channel. In the original coupled-channel amplitude, there is an $I=1$ quasi-bound state, which induces the bound state in the FCA. As discussed in Sec.~\ref{subsec:potential}, however, the energy dependence of the $DN$ potential in the variational approach is fixed at the energy of the $\Lambda_{c}^{*}$ in the $I=0$ channel. This reduces the strength of the $I=1$ amplitude, and the two-body quasi-bound state is not generated in the effective potential. Since the total spin $J=1$ channel has larger fraction of the $I=1$ $DN$ amplitude, this difference is enhanced and results in different three-body results. 

In fact, we may artificially adjust the condition~\eqref{eq:strength} to generate a quasi-bound state in the $I=1$ channel in the variational approach. By setting the strength of the $DN$ interaction at $W\sim 2766\text{ MeV}$ in the $I=1$ channel, a quasi-bound state is generated in the $I=1$ $DN$ channel. In this case, the energy dependence of the $DN$ interaction is fixed at each isospin channel, and the strength of the $DN$ attraction is  increased in the $I=1$ channel. By performing the three-body calculation, we find that the binding energy in the $J=0$ quasi-bound state are increased by 10-50 MeV, depending on the $NN$ interaction employed. This is because of the increase of the attraction, and the binding energy appears to be closer to the FCA result. In the $J=1$ sector, only the Minnesota potential supports a bound state with $B=214$ MeV, while no state is found below the $\Lambda_{c}^{*}N$ threshold with the other two $NN$ interactions. Given the uncertainty in the choice of the $NN$ interaction, the present result does not strongly support the existence of the quasi-bound state in the $J=1$ sector. In order to pin down the $J=1$ quasi-bound state, it is necessary to accumulate the experimental information of the $DN$ $I=1$ scattering amplitude, or the information on the negative parity $\Sigma_{c}^{*}$ resonance.


In addition, we should also remember that the two approaches employ different approximations. In the FCA, the dynamics of the nucleons is not solved explicitly, while the imaginary part of the $DN$ potential is not taken into account in the variational approach. In both cases, explicit $\pi Y_{c} N$ dynamics is approximated at different levels (see the discussion in Ref.~\cite{Bayar:2012rk}), whereas its importance has been pointed out in the strangeness sector~\cite{Ikeda:2008ub}. These effects can also be responsible for the difference of the results in the two approaches.

\subsection{Comparison with $\bar{K}NN$ results}\label{subsec:KNN}

It is instructive to compare the $DNN$ quasi-bound state with the corresponding $\bar{K}NN$ state in Ref.~\cite{Dote:2008hw}. In both cases, we obtain a quasi-bound state, but the $DNN$ system has a larger binding energy and a narrower width. This is in parallel with the properties of the $DN$ and $\bar{K}N$ two-body quasi-bound states, and they are closely related through the $DN$ and $\bar{K}N$ interactions.

As discussed in Sec.~\ref{subsec:amplitude}, the $D$ meson can be more strongly bound in a nucleus than $\bar{K}$ meson by two reasons. On one hand, the coupling itself is stronger, and on the other hand, the heavier mass of the $D$ meson is advantageous to increase the binding. So, we can consider two hypothetical variants between the $DNN$ system ($B\sim 230$ MeV) and $\bar{K}NN$ system ($B\sim 30$ MeV)\footnote{Here we also set the strength of the $\bar{K}N$ potential at the energy of the $\Lambda^{*}$ for comparison with the $DNN$ calculation.}; case I: kinematics of the $DNN$ system with the $\bar{K}N$ potential ($m=m_{D}, V=V_{\bar{K}}$), and case II: kinematics of the $\bar{K}NN$ system with the $DN$ potential ($m=m_{\bar{K}}, V=V_{D}$). The result of the variational calculation shows that $B\sim 40$ MeV for case I and $B\sim 190$ MeV for case II. As summarized in Table~\ref{tab:DKcomparison}, the suppression of the kinetic energy by the heavy $D$ mass is more important for the strong binding of the $DNN$ system. One should  note that in the present case, the strength of the two-body interaction is fixed at the energy of the two-body quasi-bound state. Since the $DN$ two-body bound state locates 200 MeV below the $DN$ threshold, the strength of the potential is reduced, as seen in Fig.~\ref{fig:amplitude}. Thus, in the present prescription, the attractive strength of the $DN$ potential is not very much different from the $\bar{K}N$ one, and the result of case I does not very much deviate from the $\bar{K}NN$ quasi-bound state.

\begin{table}[tdp]
\caption{Binding energies of the three-body bound state in $J=0$ channel measured from the three-body threshold with different meson mass and different meson-nucleon potential.}
\begin{center}
\begin{ruledtabular}
\begin{tabular}{lll}
                & $m=m_{\bar{K}}$ & $m=m_{D}$  \\
\hline
$V=V_{\bar{K}}$ & $\sim 30$ MeV   & $\sim 190$ MeV \\
$V=V_{D}$       & $\sim 40$ MeV   & $\sim 230$ MeV \\
\end{tabular}
\end{ruledtabular}
\end{center}
\label{tab:DKcomparison}
\end{table}%

The narrow width of the $DNN$ system is a consequence of the narrow width of the $\Lambda_{c}^{*}(2595)$. This is partly because of the small transition coupling which is suppressed by the exchange of the heavy flavor, but the main reason is the suppression of the phase space due to the large binding energy. In this sense, the heaviness of the $D$ meson is essential to realize the deep and narrow $DNN$ quasi-bound state.

\subsection{Possible experiments to produce the $DNN$ state}\label{subsec:experiment}

The very narrow width of the $DNN$ system is qualitatively different to the $\bar{K} NN$ one where the width was so large as to make its experimental observation unfeasible. In the present case there is a clear situation and there are no problems in principle for the observation of the state. In the FCA calculation, we observe that the two-nucleon absorption width is larger than the three-body decay width. This indicates that the $DNN$ quasi-bound state can be more easily seen in the two-baryon final states such as $\Lambda_{c}N$. The findings of the present work should stimulate efforts to find suitable reactions where this state could be found.

As a suggestion in this direction we can think of the $\bar{p}{~^3}\text{He} \rightarrow \bar{D}^0 D^0pn\to \bar{D}^{0} [DNN]$ reaction, which could be done by FAIR at GSI. With a $\bar{p}$ beam of $15~\text{GeV}/c$ there is plenty of energy available for this reaction and the momentum mismatch of the $D^0$ with the spectator nucleons of the $^3$He can be of the order of $550~\text{MeV}/c$, equivalent to an energy of $80~$MeV for the $D$, small compared with the scale of the binding $(\gtrsim 200$ MeV). With an estimate of $\sigma \simeq 10-20~$nb for $\bar{p}p \rightarrow \bar{D}^0 D^0$ production \cite{Kaidalov:1994mda,Khodjamirian:2011} one would expect several thousand events per day for the background of the proposed reaction \cite{Wiedner:2011mf}. A narrow peak could be visible on top of this background corresponding to the $DNN$ bound state formation.

Another possibility is the high-energy $\pi$ induced reaction. An analogous reaction is $\pi^{-} d\to D^{-}D^{+}np \to D^{-} [DNN]$ where the relevant elementary process is $\pi^{-}N\to D^{+}D^{-}N$. Since the $DN$ pair in the $DNN$ system is strongly clustering as the $\Lambda_{c}^{*}$, the reaction $\pi^{-} d\to D^{-}\Lambda_{c}^{*}n \to D^{-} [DNN]$ is also another candidate. The elementary reaction $\pi^{-}p\to D^{-}\Lambda_{c}^{*}$ is  relevant in this case. Such reactions may be realized in the high-momentum beamline project at J-PARC.

A different strategy is to look for the formation of the quasi-bound state in the heavy ion collisions. It has been shown that the hadronic molecular states with charm quark are abundantly produced at RHIC and LHC~\cite{Cho:2010db,Cho:2011ew}. Although a deeply bound $DNN$ state has smaller production yield, it can also be produced \textit{via} coalescence of the $\Lambda_{c}^{*}N$ with much smaller binding. A peak structure of the $DNN$ state may be seen, for instance, in the invariant mass spectrum of the $\Lambda_{c} \pi^{-} p$ or $\Lambda_{c}p$ final state.

\section{Conclusions}
  
We have studied the $DNN$ system with $I=1/2$ and have found that the system is bound and rather stable, with a width of about 20-40 MeV. We obtained a clear signal of the quasi-bound state for the total spin $J=0$ channel around 3500 MeV. 

We have used two methods for the evaluation of the quasi-bound state. The first one used the fixed center approximation for the Faddeev equations and the second one employs the variational approach with hadronic potentials in coordinate space. The $DN$ interaction was constructed in the field theoretical method with channel couplings and a unitary approach dynamically generates the $\Lambda_{c}^{*}(2595)$ resonance as a $DN$ quasi-bound state. 

In both cases, we have found a bound state with an energy around 3500 MeV in the $J=0$ channel. This corresponds to 250 MeV binding from the $DNN$ threshold. The $J=1$ channel is more subtle, and the precise $DN$ amplitude in the $I=1$ channel is important for a robust prediction in this channel. The mesonic decay width of the quasi-bound state turned out to be less than 40 MeV. In addition, the $D$ absorption on two nucleons was evaluated in the FCA formalism using a novel method. Although the absorption process adds several tens of MeV to the width, the total width is still much smaller than the binding energy. It is found that the $DN$ pair in $I=0$ channel in the $DNN$ system resembles the wave function of the $\Lambda_{c}(2595)$ state in vacuum. Thus, the $DNN$ state found here can be interpreted as a quasi-bound state of $\Lambda_{c}(2595)$ and a nucleon.

The small width of the $DNN$ quasi-bound state is advantageous for the experimental identification. The search for the $DNN$ quasi-bound state can be done by $\bar{p}$ induced reaction at FAIR, $\pi^{-}$ induced reaction at J-PARC, and relativistic heavy ion collisions at RHIC and LHC.

\section{Acknowledgments}
This work is partly supported by  projects FIS2006-03438 from the Ministerio de Ciencia e Innovaci\'on (Spain), FEDER funds and by the Generalitat Valenciana in the program Prometeo/2009/090.
This research is part of the European
 Community-Research Infrastructure Integrating Activity ``Study of
 Strongly Interacting Matter'' (acronym HadronPhysics2, Grant
 Agreement n. 227431) 
 under the Seventh Framework Programme of EU. 
T.H. thanks the support from the Global Center of Excellence Program by MEXT, Japan, through the Nanoscience and Quantum Physics Project of the Tokyo Institute of Technology. 
This work is partly supported by the Grant-in-Aid for Scientific Research from 
MEXT and JSPS (Nos.\
  24105702
  and 24740152).

\appendix

\section{$NN$ interactions}\label{sec:NNint}

Here we summarize the properties of the $NN$ potentials used in this study with variational calculation. We have examined three kinds of $NN$ interactions. Because we work in the isospin symmetric limit, the Coulomb interaction is not included in all cases.

Hasegawa-Nagata No.1 potential~\cite{PTP45.1786} has a three-range gaussian form with no odd force. The repulsive core is as high as 1 GeV. Because the potential was originally introduced for the RGM study, when applied to the two-nucleon systems, the attraction is too strong to generate a bound state in $^{1}S_{0}$ channel and to overestimate the deuteron binding energy. In this study, we have reduced the strength of the long-range term (middle-range term) by factor 0.25 (0.95) and call it HN1R potential. The HN1R potential has no bound state in $^{1}S_{0}$ channel and reproduces the $NN$ phase shift data, as shown below.

Minnesota force~\cite{Thompson:1977zz} is expressed by the sum of two gaussians, with relatively soft repulsive core. The parameters were chosen so as to reproduce the scattering lengths and the effective ranges of the $NN$ scattering. As a consequence, the deuteron is bound only with the $s$-wave component, so the tensor force is considered to be renormalized in the central part. We set the parameter $u=1$ so that there is no odd force, unless otherwise stated.

Argonne v18 potential~\cite{Wiringa:1994wb} is one of the realistic nuclear forces with strong repulsive core. As in Ref.~\cite{Dote:2008hw}, we used the gaussian-fitted version of the potential with central, spin-spin, and $L^{2}$ terms. Since the description of the deuteron requires the $d$-wave mixing
which is beyond the present model wave function, we only consider the $S=0$ channel with the Av18 potential.

\begin{figure}[tbp]
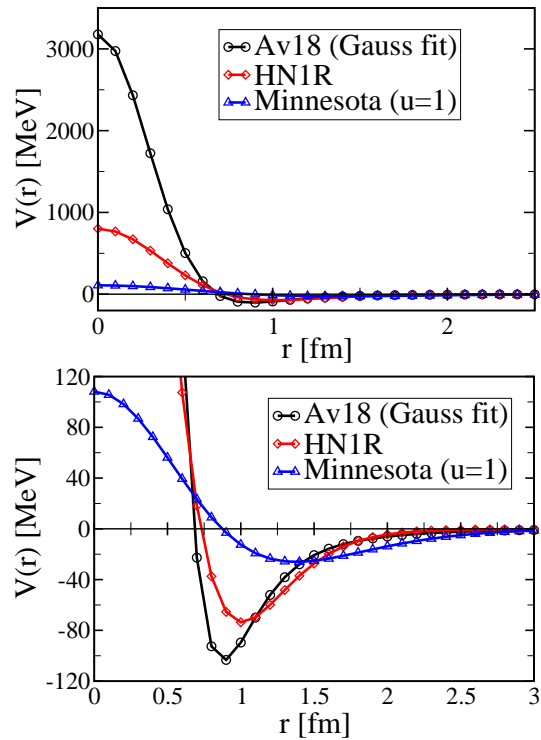

    \centering
    \includegraphics[width=7cm,clip]{NNpot1_v2.eps}
    \includegraphics[width=7cm,clip]{NNpot2_v3.eps}
    \caption{\label{fig:potentialform}
    (Color online) Coordinate space $NN$ potentials in the $^{1}S_{0}$ channel.}
\end{figure}

\begin{figure}[tbp]
    \centering
    \includegraphics[width=7cm,clip]{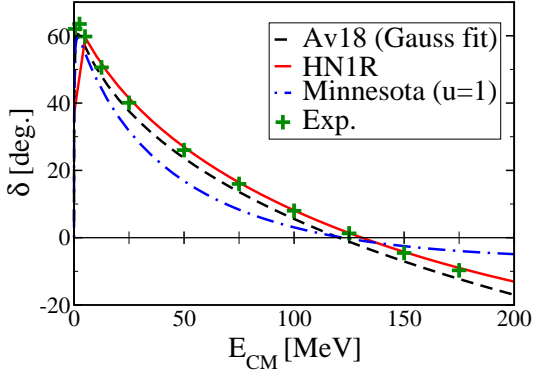}
    \caption{\label{fig:phaseshift}
    (Color online) $NN$ phase shifts in the $^{1}S_{0}$ channel calculated by the $NN$ potentials.}
\end{figure}

In Fig.~\ref{fig:potentialform}, we show the spatial form of the potentials in the $^{1}S_{0}$ channel. The phase shifts of the $NN$ scattering in the $^{1}S_{0}$ channel are shown in Fig.~\ref{fig:phaseshift} in comparison with experimental data.

\section{Derivation of the three-body amplitude in the charge basis}\label{sec:chargebasis}

In this Appendix, we derive Eqs.~\eqref{eq:TJ0} and \eqref{eq:TJ1} in the approach of Ref.~\cite{Bayar:2012rk} by applying the following strategy. We evaluate first the $D^0 p p \rightarrow D^0 p p $ amplitude considering charge exchange processes in the rescattering of the $D$ meson. The amplitude will contains total isospin $I_{\text{tot}}=1/2$ and $I_{\text{tot}}=3/2$. Since we only want the $I_{\text{tot}}=1/2$, we evaluate the scattering amplitude for $I_{\text{tot}}=3/2$ in addition, taking the $D^+ p p \rightarrow D^+ p p $ amplitude, and from a linear combination of the two we obtain the $I_{\text{tot}}=1/2$ amplitude. This strategy was found most practical in Ref.~\cite{Bayar:2012rk}.

For the $D^0 p p \rightarrow D^0 p p $ we define three components of the three-body scattering amplitude; 

a) $T_p$, which are called partition functions, which contains all diagrams that begin with a $D^0$ collision  with the first proton of the $pp$ system and finish with $D^0 p p$,

b) $T_{ex}^{(p)}$, which contains all the diagrams that begin with a $D^+ p$ collision on a $n p$ system and finish with $D^0 p p$, and

c) $T_{ex}^{(n)}$, which contains all the diagrams that begin with a $D^+n$ collision on a $np$ system and finish with 
$D^0 p p $.

\begin{figure}[tbp]
\begin{center}
\includegraphics[width=0.5\textwidth] {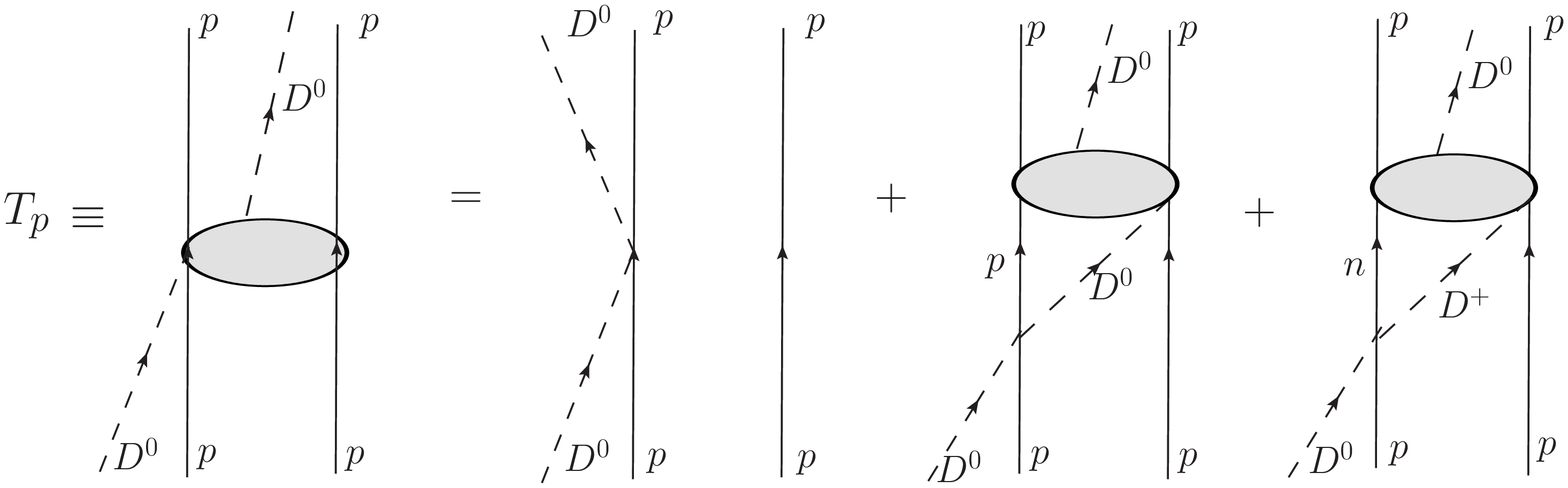}\\
\includegraphics[width=0.3\textwidth] {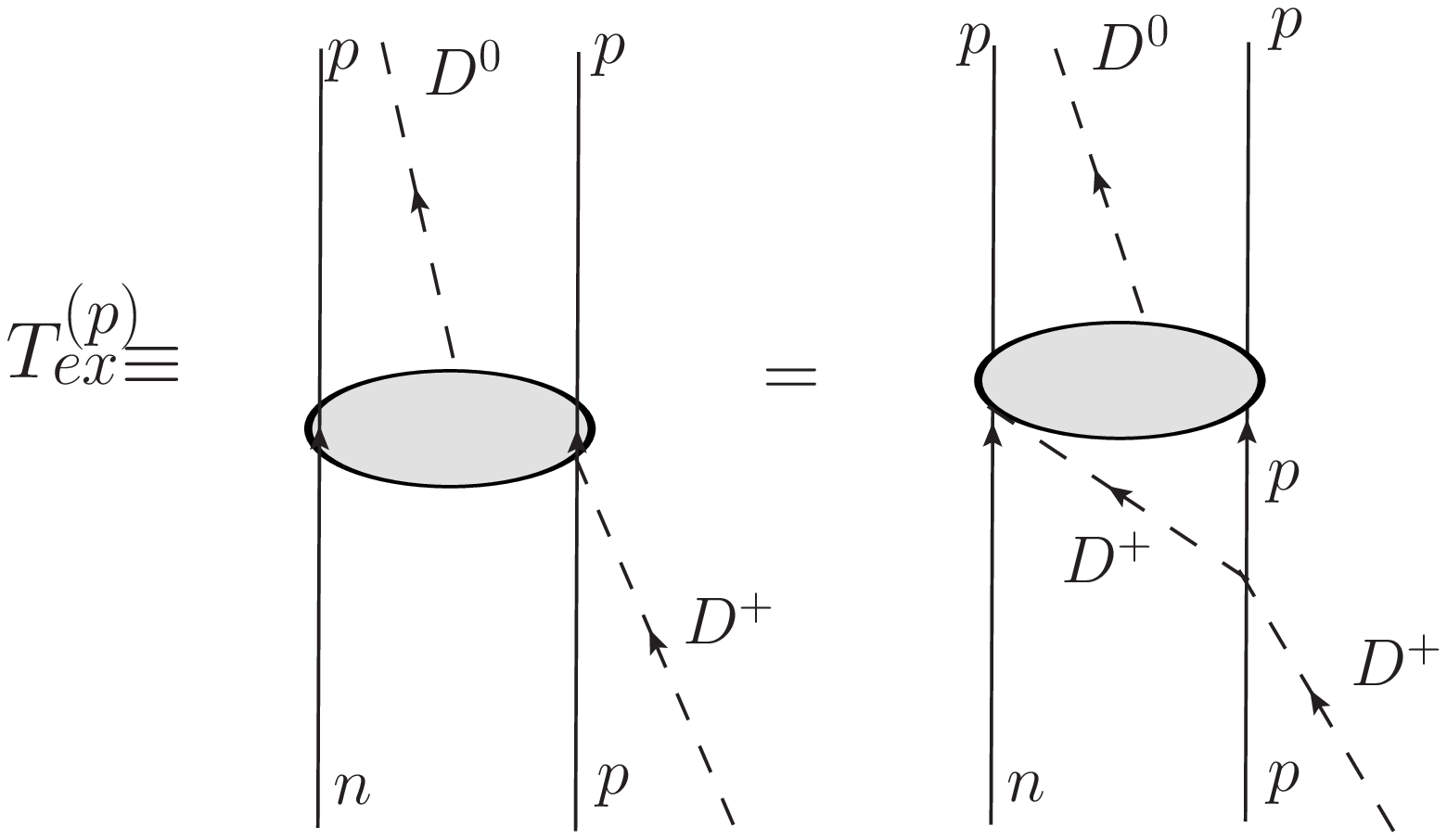}\\
\includegraphics[width=0.5\textwidth] {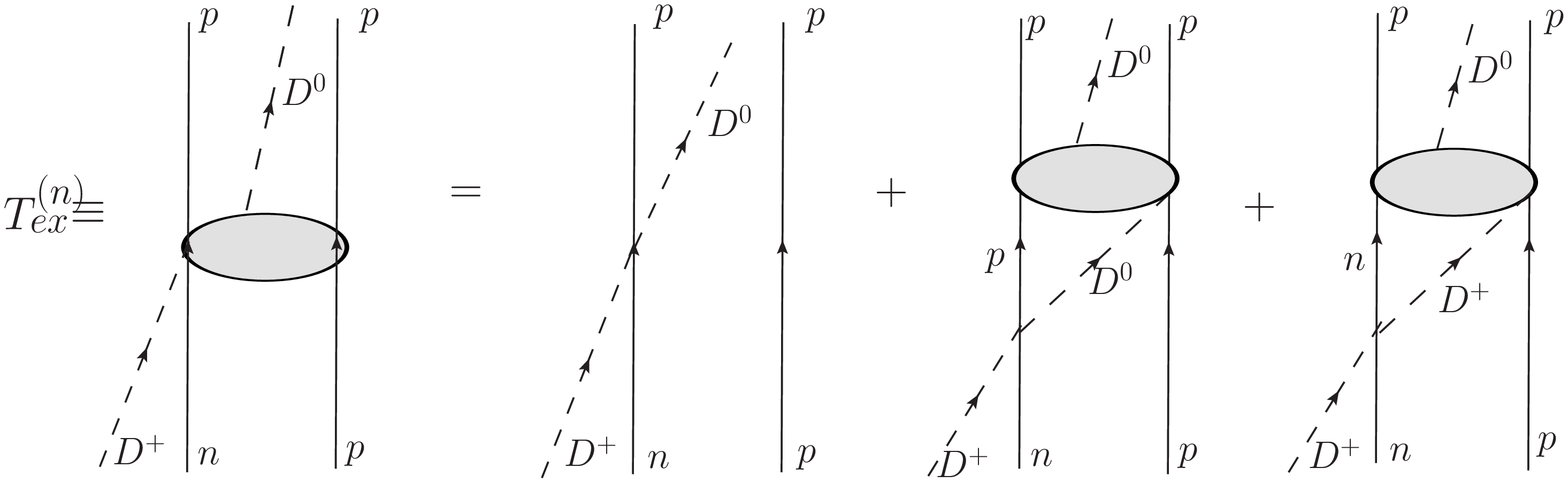}
\caption{Diagrammatic representations of the partition functions for the $D^0 p p \rightarrow D^0 p p $.}
\label{fig:tpp22}
\end{center}
\end{figure}

These amplitudes fulfill a set of coupled equations 
\begin{eqnarray}
T_{p}&=&t_{p}+t_{p}G_0T_{p}+t_{ex}G_0T_{ex}^{(p)}\nonumber\\
T_{ex}^{(p)}&=&t_{0}^{(p)}G_0T_{ex}^{(n)}\nonumber\\
T_{ex}^{(n)}&=&t_{ex}+t_{ex}G_0T_{p}+t_{0}^{(n)}G_0T_{ex}^{(p)}
\label{Eq:tptexp}
\end{eqnarray} 
where the two-body amplitudes are given as
$t_p=t_{D^0 p , D^0 p}$, $t_{ex}=t_{D^0 p , D^+ n}$, $t_{0}^{(p)}=t_{D^+ p , D^+ p}$, and $t_{0}^{(n)}=t_{D^+ n , D^+ n}$. The set of equations~\eqref{Eq:tptexp} are diagrammatically represented in Fig.~\ref{fig:tpp22}.

By taking into account the phase convention $|D^0\rangle =-|1/2,-1/2\rangle $ in the isospin basis, we can write all the former elementary amplitudes in terms of $I=0,~1$ ($t^{(0)},~t^{(1)}$) for the $DN$ system, and we find
\begin{eqnarray}
t_{p}&=&\frac{1}{2}(t^{(0)}+t^{(1)})\nonumber\\
t_{ex}&=&\frac{1}{2}(t^{(0)}-t^{(1)})\nonumber\\
t_{0}^{(p)}&=&t^{(1)}\nonumber\\
t_{0}^{(n)}&=&\frac{1}{2}(t^{(0)}+t^{(1)}).
\label{Eq:tmatpex}
\end{eqnarray} 
Eliminating $T_{ex}^{(p)}$ and $T_{ex}^{(n)}$ in Eq. (\ref{Eq:tptexp}) we obtain 
\begin{equation}
T_{p}=\frac{t_{p}(1-t_{0}^{(n)}G_0t_{0}^{(p)}G_0)+t_{ex}^2G_0t_{0}^{(p)}G_0}{(1-t_{p}G_0)(1-t_{0}^{(n)}G_0t_{0}^{(p)}G_0)-t_{ex}^2t_{0}^{(p)}G_0^3}
\label{Eq:tpp}
\end{equation}
which in isospin basis can be simplified to
\begin{equation}
T_{p}=\frac{\frac{1}{2}(t^{(0)}+t^{(1)})-t^{(0)}t^{(1)^2}G_0^2}{(1-G_0t^{(1)})(1+\frac{1}{2}(t^{(1)}-t^{(0)})G_0-G_0^2t^{(0)}t^{(1)})}
\nonumber
\end{equation}
The total $D^0 p p \rightarrow D^0 p p $ amplitude would be $2T_{p}$ accounting for the first interaction of the $D^0$ with either of the protons.

\begin{figure}[tbp]
\includegraphics[width=0.5\textwidth]{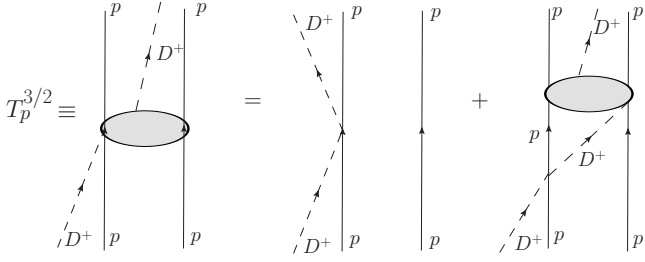}
\caption{Diagrammatic representation of the partition function for I=3/2.}\label{tpp3b2}
\end{figure}

Now we take into account that in the basis of $|I_{tot},I_{3,tot}\rangle$
\begin{equation}
|D^0 p p\rangle =-(\dfrac{1}{\sqrt{3}}|3/2,1/2\rangle
+\sqrt{\dfrac{2}{3}}|1/2,1/2\rangle )
\end{equation}
and thus
 \begin{equation}
\langle 1/2|T|1/2\rangle =\dfrac{3}{2}(\langle D^0 p p|T|D^0 p p\rangle -\dfrac{1}{3}\langle 3/2|T|3/2\rangle)
\label{Eq:t1b2}
\end{equation}
The $\langle 3/2|T|3/2\rangle$ amplitude is particularly easy to obtain. In this case we take the $D^+ p p \rightarrow D^+ p p $ transition and diagrammatically we have the mechanism of Fig. \ref{tpp3b2} for the only partition function $T_p^{(3/2)}$.

Hence
\begin{equation}
T_p^{(3/2)}=t_{0}^{(p)}+t_{0}^{(p)}G_0T_p^{(3/2)}
\end{equation}
and the total $T^{(3/2)}$ amplitude will be  $2T_p^{(3/2)}$, accounting for the $D^+$ interacting first also with the second nucleon. We have 
\begin{equation}
T_p^{(3/2)}=\dfrac{t_{0}^{(p)}}{1-G_0t_{0}^{(p)}}=\dfrac{t^{(1)}}{1-G_0t^{(1)}}.
\end{equation}
We can now use Eq. (\ref {Eq:t1b2}) and find for the total amplitude (including the factor two for first interaction with either proton)
\begin{eqnarray}
T^{(1/2)}&=&3T_p-\dfrac{t^{(1)}}{1-G_0t^{(1)}}\nonumber\\
&=&\dfrac{\dfrac{3}{2}t^{(0)}+\dfrac{1}{2}t^{(1)}-\dfrac{1}{2}t^{(1)}(t^{(1)}-t^{(0)})G_0-2t^{(0)}t^{(1)^2}G_0^2}{(1-G_0t^{(1)})(1+\frac{1}{2}(t^{(1)}-t^{(0)})G_0-G_0^2t^{(0)}t^{(1)})}
\nonumber
\end{eqnarray}
which can be simplified dividing the numerator by $(1-G_0t^{(1)})$ with the final result
\begin{eqnarray}
T^{(1/2)}&=&\dfrac{\dfrac{3}{2}t^{(0)}+\dfrac{1}{2}t^{(1)}+2G_0t^{(0)}t^{(1)}}{1+\frac{1}{2}(t^{(1)}-t^{(0)})G_0-G_0^2t^{(0)}t^{(1)}}.
\nonumber
\end{eqnarray}
This corresponds to Eq.~\eqref{eq:TJ0}.

The case of $S_{NN}=1$ $(I_{NN}=0)$, that we also study here, can be done in  a similar way, but this was done in \cite{Kamalov:2000iy} for $K^- d$ at rest and in \cite{Oset:2012gi} for the $K^- d$ interaction below threshold. We quote here the formula that was obtained in  \cite{Oset:2012gi} which we use here too.
\begin{eqnarray}
T_{D^0 d}&=&\dfrac{\dfrac{1}{2}t^{(0)}+\dfrac{3}{2}t^{(1)}+2G_0t^{(0)}t^{(1)}}{1-\frac{1}{2}(t^{(1)}-t^{(0)})G_0-G_0^2t^{(0)}t^{(1)}}.
\nonumber
\end{eqnarray}
This corresponds to Eq.~\eqref{eq:TJ1}.

\end{document}